\documentclass[aps,pro,superscriptaddress,twocolumn,floatfix]{revtex4-1}
\usepackage{amsfonts}
\usepackage{amsmath}
\usepackage{amssymb}
\usepackage{braket}
\usepackage{hyperref}
\usepackage{graphicx,dsfont}

\def\beq{\begin{equation}}
\def\eeq{\end{equation}}

\def\f{\frac}
\def\l{\left}
\def\r{\right}
\def\p{^+}
\def\m{^-}
\def\z{^z}

\def\d{^\dagger}

\def\up{\uparrow}
\def\down{\downarrow}
\def\k{\mathbf k}

\def\bx{\begin{pmatrix}}
\def\ex{\end{pmatrix}}
\def\n{^{\;}}

\begin{document}

\title{Integrable Model of Topological SO(5) Superfluidity}

\author{Will J. Holdhusen}
\affiliation{Department of Physics, Indiana University, Bloomington, Indiana 47405, USA}

\author{Sergio Lerma-Hern\'andez}
\affiliation{Facultad de F\'{i}sica, Universidad Veracruzana, Circuito Aguirre Beltr\'an s/n, Xalapa, Veracruz 91000, Mexico}

\author{Jorge Dukelsky}
\affiliation{Instituto de Estructura de la Materia, CSIC, Serrano 123, 28006 Madrid, Spain}

\author{Gerardo Ortiz}
\affiliation{Department of Physics, Indiana University, Bloomington, Indiana 47405, USA}

\begin{abstract}
Assisted by general symmetry arguments and a many-body invariant, we introduce 
a phase of matter that constitutes a topological SO(5) superfluid. Key to this 
finding is the realization of an exactly solvable model that displays some 
similarities with a minimal model of superfluid $^3$He. We study its quantum phase 
diagram and correlations, and find exotic superfluid as well as metallic phases
in the repulsive sector. At the critical point separating trivial and non-trivial 
superfluid phases, our Hamiltonian reduces to the globally SO(5)-symmetric Gaudin model 
with a degenerate ground manifold that includes quartet states. Most importantly, 
the exact solution permits uncovering of an interesting non-pair-breaking mechanism 
for superfluids subject to external magnetic fields. Non-integrable modifications of 
our model lead to a strong-coupling limit of our metallic phase with a ground state manifold 
that shows an extensive entropy. 
\end{abstract}

\maketitle

{\it Introduction} --- 
Exactly-solvable models of quantum many-body systems are theoretical constructions key 
to uncover physical mechanisms or effects resulting from competing interactions 
\cite{Sutherland_2004, Gaudin_1995, ortiz05}. 
The case of spin-1/2 particles with SO(5)-symmetric $p$-wave interactions is particularly 
compelling because it can give rise to non-trivial spin-triplet Cooper-pair topological phases 
with no equivalent in SU(2)-symmetric couplings. For instance, it is  well-known that the 
emergent SO(3)$_{\bf L}\otimes$SO(3)$_{\bf S}\otimes$U(1) symmetry in liquid $^3$He, 
contained in SO(5), is responsible for topological classification of the defects of 
its exotic superfluid phases \cite{Volovik_2009}. Similar mechanisms 
could be at play in unconventional uranium-based metallic ferromagnetic superconductors, 
where strong external magnetic fields can even revive superconductivity \cite{AnnedeVisser_2020}. 
A theoretical understanding of these mechanisms is therefore a prerequisite to engineering materials or 
synthetic matter with exotic magnetic superfluid behavior \cite{Galitskii_2019}. 

SO(5)-symmetric models have a long history in nuclear physics
as a description of isovector (isospin 1) pairing between
protons and neutrons.
The earliest version of an integrable model consisting of a unique 
SO(5) algebra, describing a proton-neutron system, was presented in Refs.
\cite{hecht65, ginocchio65}. The generalization of the
exact solution to many
SO(5) copies or, equivalently, 
to many non-degenerate single particle 
orbitals arose as an extension of the Richardson-Gaudin (RG) models 
\cite{dukelsky01, dukelsky04_2} to rank 2 algebras 
\cite{links02, dukelsky06}.  
In condensed matter physics, the systems closest to 
admitting an SO(5)-symmetric representation are arguably superfluid $^3$He 
\cite{leggett75,Hasegawa1979,Zhang_2002,murakami99} and non-$p$-wave systems 
\cite{wu03, demler04, wu05},
but, as far as we know, there are no corresponding integrable 
interacting models.

In this work, we study the quantum phase diagram of the 
fermionic ($c^{\;}_{\k\sigma}, c_{\k\sigma}\d$) SO(5) Hamiltonian expressed in momentum ($\k$) space as
\begin{eqnarray} \hspace*{-0.7cm}
H &=& \sum_{\k, \sigma=\uparrow,\downarrow}\epsilon_\k 
c_{\k\sigma}\d c^{\;}_{\k\sigma}
- \sum_{\k,\k'}\Delta_{\k\k'}^{\;} \left(
\vec T_\k\p \cdot \vec T_{\k'}\m
+ \vec T_\k \m \cdot \vec T_{\k'}\p\right) \nonumber \\
&-& \sum_{\k,\k'}W_{\k\k'}^{\;} \vec S_\k^{\;} \cdot \vec S_{\k'}^{\;}
-  \sum_{\k,\k'}V_{\k\k'}^{\;} N_\k^{\;} N_{\k'}^{\;}- h \sum_\k S_\k^z.
\label{eqn:hamiltonian}
\end{eqnarray}
The operator
$T_{\mu \k}\p$ ($\vec T_\k\p = (T_{-1\k}\p, T_{0\k}\p, T_{1\k}\p)$) creates a spin-triplet fermion pair $(\k,-\k)$ with
spin-projection $\mu=\pm 1, 0$. 
Magnetic Heisenberg ($\vec S_\k \cdot \vec S_{\k'}$)
and density-density ($N_\k N_{\k'}$, where
$N_\k$ counts all spinful fermions with momenta $\pm \k$)
interactions
complete the minimal set required
to close an SO(5) algebra, with a fermionic
representation \cite{SM}
\begin{eqnarray}
&T_{\mu \k}\m &= \frac{(-1)^{\frac{\mu(\mu+1)}{2}}}{(2\delta_{\mu,\pm 1}+\sqrt{2}\delta_{\mu, 0})} \sum_{\sigma,\sigma'}c_{-\k\sigma}\n\l(i \sigma^\mu\sigma^y\r)_{\sigma\sigma'}c_{\k\sigma'}\n ,
\nonumber \\
&S_{\k}^\mu &= \frac{1}{2}\sum_{\sigma,\sigma'} (
c_{\k\sigma}\d \sigma_{\sigma\sigma'}^\mu
c_{\k\sigma'}\n + 
c_{-\k\sigma}\d \sigma_{\sigma\sigma'}^\mu
c_{-\k\sigma'}\n ),
\label{eqn:so5}
\\
&N_\k\n &=\sum_{\sigma} ( c_{\k\sigma}\d c_{\k\sigma}\n
+ c_{-\k\sigma}\d c_{-\k\sigma}\n), \nonumber
\end{eqnarray}
where $\sigma^{\pm}=\sigma^x\pm i \sigma^y$, $\sigma^0=\sigma^z$ are Pauli matrices and 
$T_{\mu\k}\p=(T_{\mu \k}\m)\d$. 
For an appropriate choice of (separable) interactions $\Delta_{\k\k'} = W_{\k\k'}=4V_{\k\k'}$, 
Hamiltonian \eqref{eqn:hamiltonian} is exactly-solvable independently of spatial dimensionality.

In the attractive pairing
sector, the model displays trivial and 
non-trivial topological superfluid phases, separated by a critical point that is 
globally SO(5)-symmetric. At this point the ground state manifold is macroscopically 
degenerate with pair and quartet correlations.
The application of a magnetic field $h$ \cite{Hasegawa1980}
leads to a remarkable magnetized superfluid,
where spin-triplet pairs transition, without pair breaking, between different spin projections 
as in the B to A first-order transition in superfluid $^3$He. This mechanism is absent in SU(2) pairing 
models. Finally, the repulsive sector shows a metallic phase whose strong-coupling limit is adiabatically
connected to a flat band model with 
exponentially-degenerate ground states, giving an extensive entropy
similar to holographic models of strange metals \cite{sachdev10}.

{\it Exactly-solvable SO(5) model} --- 
RG integrable models are defined by a set of integrals 
of motion $R_\k$ fulfilling the integrability condition 
$[R_\k, R_{\k'}] = 0$, 
such that their linear combination realizes a Hamiltonian as \eqref{eqn:ham_iom}. 
The exact eigenspectrum of the integrals of motion, and corresponding Hamiltonian, 
may be found with algebraic complexity by solving the RG ansatz equations.

These models may be 
formulated in terms of a generalized Gaudin algebra
\cite{ortiz05, H_2007, Errea_2009, subero09}.
Starting from the rational SO(5) RG integrals of motion
\cite{dukelsky06}, we form the set 
\beq
R_\k = 
\l(1+\frac{\Delta}{2}\r)N_\k^- +\frac{\Delta}{2} S_\k\z
+ q\sum_{\k'\neq \k}Z_{\k\k'}
\vec{\mathcal T}_{\k}\cdot
\vec{\mathcal T}_{\k'} ,
\label{eqn:iom}
\eeq
where 
$N_\k\m = N_\k/2-1$,
and $\vec{\mathcal T}_\k^{\;} \cdot \vec{\mathcal T}_{\k'}^{\;}$
is the SO(5) Gaudin interaction
$
\vec{\mathcal T}_\k^{\;} \cdot \vec{\mathcal T}_{\k'}^{\;}
=
\vec T_\k\p \cdot \vec T_{\k'}\m
+ \vec T_\k\m \cdot \vec T_{\k'}\p
+ \vec S_\k \cdot \vec S_{\k'}
+ N_\k^- N_{\k'}^-$.
The function
$Z_{\k\k'} =Z(\eta_\k,\eta_{\k'})= \frac{\eta_\k \eta_{\k'}}{\eta_\k - \eta_{\k'}}$, $Z_{\k\k'}=-Z_{\k'\k}$,
is a particular case of a more general function interpolating between hyperbolic and trigonometric SU(2) RG models \cite{richardson02, ortiz05, rombouts10}.
The parameters $\Delta$, $q$, and $\eta_{\k}$ are arbitrary real numbers with the restriction that $\eta_\k \neq \eta_{\k'}$ for $\k \neq \k'$ to avoid singularities.

In Eq. \eqref{eqn:iom} and for the remainder of this paper,
sums are taken over momenta with $k_x > 0$ to avoid double counting.
Each pair of momenta $\pm \k$ labels a level with a 
corresponding irreducible representation (irrep)
of SO(5) characterized by seniority
$\nu_\k$ and reduced
spin $s_\k$ quantum numbers. The $l$ levels correspond to a lattice with 
$L = 2l$ sites, since each level incorporates two modes 
in $\k$-space.

Eigenvalues and eigenvectors of the integrals of motion are
determined by two sets of spectral parameters: pairons $e_\alpha$, $\alpha = 1, \dots , N_e$, and
wavefunction parameters 
$\omega_\beta$, $\beta = 1, \dots , N_\omega$, 
that are roots of the two sets of RG (Bethe) equations
\begin{align}
\begin{split}\hspace*{-0.2cm}
-\f{1}{q} =&
2 \! \sum_{\alpha'\neq \alpha} \! Z_{\alpha'\alpha}
-\sum_{\beta}  Z_{\beta\alpha}
+ \sum_\k \! \l(\f{\nu_\k}{2}-1+s_\k\r)Z_{\k \alpha} ,
\end{split}
\label{eqn:gaudin_eqs_1}
\end{align}
and
\begin{align}
\begin{split}
-\frac{\Delta}{q} =&
-\sum_{\beta'\neq \beta}Z_{\beta'\beta}
+ \sum_\alpha Z_{\alpha\beta}
+ \sum_\k s_\k Z_{\k \beta} ,
\end{split}
\label{eqn:gaudin_eqs_2}
\end{align}
with $Z_{\alpha'\alpha}=Z(e_{\alpha'}, e_\alpha)$, 
$Z_{\beta\alpha}=Z(\omega_\beta, e_\alpha)$, 
and $Z_{\k \alpha}=Z(\eta_\k, e_\alpha)$.
The number of pairons $N_e$ is equal to the number of 
spin-1 fermion pairs and relates to the total fermion 
number as $N =2 N_e + \sum_\k \nu_\k$. The number of 
wavefunction parameters is 
$N_\omega = N_e + \sum_\k (S_\k\z + s_\k)$. We emphasize that while the 
dimension of the Hilbert space grows exponentially with the number of levels $L$ 
and $N_e$, the complexity of the exact solution 
grows only linearly, allowing exact treatment of very large systems.

Numerical solution of the RG
equations must navigate the singularities that
arise whenever the spectral parameters approach each 
other or the level parameters $\eta_\k$.
To avoid these singularities, we add modulated
imaginary parts to $\eta_\k$ while iteratively solving to a desired coupling
$q$ and then incrementally remove these to achieve physical results \cite{rombouts10}. 

In terms of the variables $e_\alpha$ and $\omega_\beta$, the integrals of motion $R_\k$ have eigenvalues
\begin{eqnarray}
r_\k &=&\frac{\nu_\k}{2} -1+qs_\k \sum_\beta Z_{\k\beta} +q\l(1-\f{\nu_\k}{2}-s_\k\r)\sum_\alpha Z_{\k\alpha} \nonumber
\\
&&-q\sum_{\k'\neq \k}\l[\l(\f{\nu_\k}{2}-1\r)\l(\f{\nu_{\k'}}{2}-1\r) +
s_\k s_{\k'}\r]Z_{\k\k'}.
\label{eqn:full_gen_eigens}
\end{eqnarray}
To obtain the corresponding eigenstates, we need
operators
\beq
\mathsf{S}_\beta\p = \sum_\k Z_{\k\beta} \, S_\k^+,
\quad
\mathsf{T}_{\mu\alpha}\p= \sum_\k Z_{\k\alpha} \, T_{\mu \k}^+ ,
\eeq
and $\overleftarrow{\mathsf I}_\alpha^+$,
defined by its action on $\mathsf{T}_{\mu\alpha}\p$:
\beq
\mathsf{T}_{\mu\alpha}\p\overleftarrow{\mathsf I}^+_{\alpha'}
= \delta_{\alpha\alpha'}\begin{cases}
\mathsf{T}_{\mu + 1\alpha}\p & \mu \le 0 \\
0 & \mu = 1
\end{cases}    .
\eeq
Then, the eigenstates can be written as \cite{SM}
\begin{align}
\begin{split}
\ket{\Psi}=\prod_{\alpha=1}^{N_e} \mathsf T_{-1\alpha}\p
\prod_{\beta=1}^{N_\omega}
\left(
\mathsf S_\beta^+ - \sum_{\alpha'=1}^{N_e} 
\overleftarrow{\mathsf I}_{\alpha'}^+
Z^*_{\alpha'\beta}
\right)\ket \Lambda ,
\end{split}
\label{eqn:wavefunction}
\end{align}
where $\ket \Lambda$ is a vacuum state 
characterized by $\nu_\k$ and $s_\k$ with $S_\k\m \ket \Lambda = T_{\mu \k}\m \ket \Lambda = 0$.
In most cases, the ground state is built from the empty ($\nu_\k = s_\k = 0$ for all $\k$)
vacuum $\ket 0$. The exception occurs
when sufficiently strong repulsive pairing couplings break pairs.

When $\epsilon_\k=\eta_\k$, $\Delta_{\k\k'}= W_{\k\k'} = 4 V_{\k\k'} = (g/L) \eta_\k \eta_{\k'}$, and 
$h=0$, 
Hamiltonian \eqref{eqn:hamiltonian} can be 
written as  a linear combination of the integrals of motion \cite{SM}
\begin{eqnarray}
H &=& \frac{2}{1 - q\sum_\k \eta_\k}\sum_\k \eta_\k R_\k  +
\text{constant}\label{eqn:ham_iom} \\
 &=&\l(1 - \frac{g}{g_c}\r)\sum_\k \eta_\k N_\k 
- \frac{g}{L} \sum_{\k,\k'}\eta_\k \eta_{\k'}
\vec{\mathcal T}_\k\cdot \vec{\mathcal T}_{\k'} + \frac{gL}{g_c^2}.
\nonumber
\end{eqnarray}
Here, we define
$qL =- g/(1 - g/g_c)$ with 
$g_c^{-1}L = \sum_\k \eta_\k$ and set $\Delta = 0$ (letting $\Delta \neq 0$ has the effect of assigning a different kinetic energy to spin-up versus spin-down fermions).
At $g = g_c$, $q$ becomes singular and $H$ reduces to the (globally) SO(5)-symmetric Gaudin model, as it is  evident from the second line in (\ref{eqn:ham_iom}).
Adding a uniform magnetic field $h$ does not break integrability and is discussed below.  

Using the eigenvalues $r_\k$ from Eq. \eqref{eqn:full_gen_eigens},
the total energy for a system of density $\rho = N/L$ is
\begin{eqnarray}
\mathcal E(N)= \frac{2 \sum_\k \eta_\k r_\k}{1-q\sum_\k \eta_\k}
\end{eqnarray}
up to a constant dependent on the vacuum state $\ket \nu$ \cite{SM}
while total energy per site (energy density) will be indicated by $e = \mathcal E(N)/L$. Other observables may also be computed from the integrals of motion using the Hellmann-Feynman theorem, i.e., momentum distribution 
$\braket{N_\k} = 2(r_\k -  q\partial r_\k/\partial q+1)$.

Hamiltonian \eqref{eqn:ham_iom} displays a particle-hole symmetry $\mathcal P$.  
Under the map 
$\mathcal P\d c_{\k\up} \mathcal P = c_{\k\down}\d$,
$\mathcal P\d c_{\k\down} \mathcal P = c_{\k\up}\d$,
$H$ transforms  as
$\mathcal P\d H(\rho, g)\mathcal P
= \alpha H\l( 2-\rho, \alpha g\r)$ up to an additive 
constant, where $\alpha^{-1} = 2g/g_c-1$. At $g=g_c$, $\alpha=1$ and $H$ is particle-hole
symmetric.

{\it Quantum phase diagram: One-dimensional case} --- To illustrate 
the physics arising from our SO(5) model we now work in one spatial dimension.
The momenta for periodic boundary conditions are
$k_j=\frac{2\pi j}{L}$, $j = -L/2, -L/2+1, \dots, L/2-1$,
which leaves isolated modes at $k=0$ and $k=-\pi$
that cannot participate in pairing
interactions and therefore do not correspond to
RG levels.
When using these boundary
conditions, we ignore all interactions on the $k=-\pi$ mode
to preserve integrability. The effect of the
ignored interactions diminishes in the
thermodynamic limit ($L\rightarrow \infty$ with $\rho$ and $g$ fixed).
To avoid this finite-size effect, the majority of our
calculations utilize antiperiodic boundary
conditions, under which all momenta
come in pairs $(+|k|, -|k|)$
corresponding to RG levels:
$k_j = \frac{\pi}{L}(2j+1)$, $j =-L/2, -L/2+1, \dots, L/2-1$. 

We linearize the dispersion close to 
the Fermi points by choosing $\eta_k=k$ and $\epsilon_k =|k|$ (in units of $\frac{1}{2}\hbar v_F$ where $\hbar$ is Planck's constant and  $v_F$ the Fermi velocity).
Because $\eta_k = - \eta_{-k}$, interaction coefficients
$\Delta_{kk'}=\eta_k\eta_{k'}$ have the antisymmetry necessary
for $p$-wave pairing: 
$\Delta_{kk'} = \Delta_{-k-k'}=-\Delta_{-k k'}$.
In coordinate space, 
the Fourier-transformed coefficients $\Delta_{ij}$ linking sites at $r_i$ and $r_j$ decay as $(r_i -r_j)^{-1}$ \cite{ortiz14, ortiz16},
and show alternating sign $(-1)^{i-j}$\cite{SM}.

We next analyze the various phases that emerge in the phase diagram of 
the SO(5) RG Hamiltonian \eqref{eqn:ham_iom}. 
\begin{figure}
\includegraphics{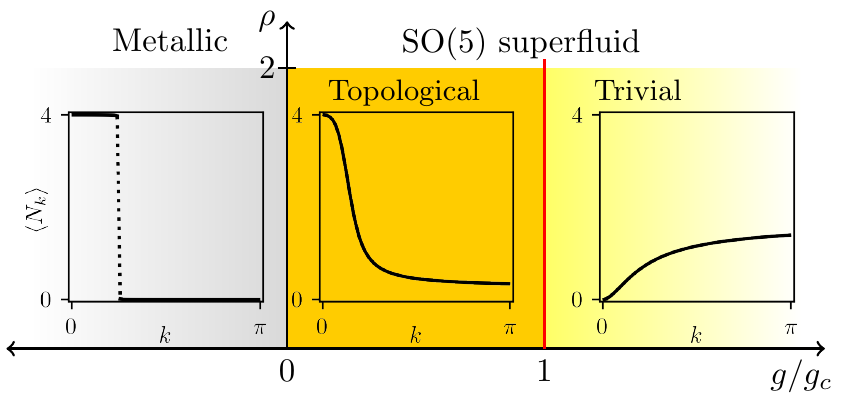}

\caption{
Quantum phase diagram of the SO(5) RG model as a 
function of density $\rho$ and coupling $g$. 
Ground-state momentum distributions 
$\braket{N_k}$, as a function of $k>0$
for $\rho=1/2$, 
are displayed at $g/g_c=-1,\frac{3}{4},\frac{3}{2}$,
that is, for the 
metallic, topological, and 
trivial paired-superfluid phases, respectively.
The metallic phase displays a discontinuity at the Fermi momentum $k_F=\pi\rho/2$, the topological superfluid phase is continuous with occupation of low-momentum modes, and the trivial superfluid vacates the 
low-momentum modes. The phase transition between superfluid phases is second-order (see Fig. \ref{fig:derivs}).
}
\label{fig:phases}
\end{figure}

\begin{figure}\hspace*{-0.4cm}
\includegraphics{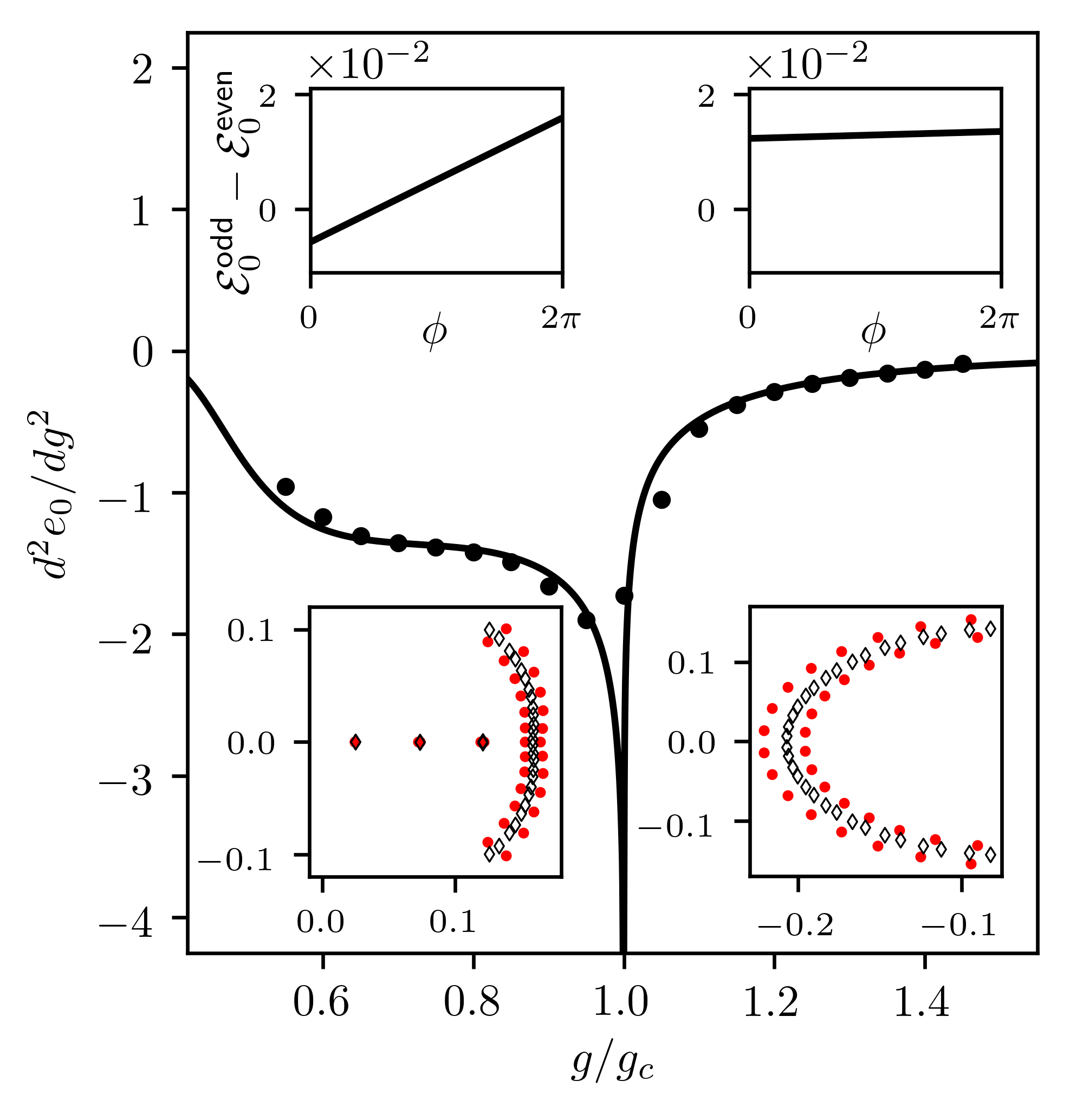}
\caption{Second derivative of the ground-state energy density $e_0$
with respect to coupling for a mean-field solution \cite{SM} with $L = 2000$ (solid line), and the exact solution
at $L = 128$, both at quarter filling ($\rho=1/2$). 
Lower insets: the variables
$e_\alpha$ (diamonds) and $\omega_\beta$ (dots) are plotted with their imaginary parts on the $y$-axis
and their real parts on the $x$-axis
at $g/g_c = 0.9$ (left) and 1.1 (right).
An animation of these spectral parameters as a function of $g$
is included in the supplemental material \cite{SM}.
Upper insets: energy difference
between odd and even sectors
is plotted as a function of boundary condition $\phi$ for
a quarter-filled system with 32 fermions at 
$g/g_c = 0.7$ (left) and $1.75$ (right).
% The parity switch $\mathcal P_N(\phi)=\text{sign}\l(
% \mathcal E_0^{\sf odd} (\phi) -
% \mathcal E_0^{\sf even} (\phi)
% \r)$ changes in the topological phase as a function of  $\phi$.
}
\label{fig:derivs}
\end{figure}

{\it Topological superfluid phase} --- For attractive couplings $g > 0$, the
ground state of Hamiltonian \eqref{eqn:ham_iom} 
is a superfluid of spin-triplet pairs.  At  the density-independent critical coupling $g=g_c$, the system 
undergoes a 
topological phase transition
with an accompanying change in occupation number 
$\braket{N_k}$ around zero momentum ($k=0$),
as seen in the inset of Fig. \ref{fig:phases}.
The system transitions from a weak-pairing topologically 
non-trivial SO(5) superfluid into a strong-pairing trivial 
superfluid gapped phase. The transition is signaled by
a divergence in $\partial^2 e_0/\partial g^2$,
the second-order derivative of the ground state energy
density.
(Figure \ref{fig:derivs} illustrates this along with distribution of spectral parameters in the complex plane).

To understand the topological nature of these superfluid phases, we need a 
many-body (bulk) topological invariant distinguishing them. Ref. 
\cite{ortiz14} introduces a fermion parity switch for spinless fermions
that distinguished $p$-wave topological phases of an SU(2) model. Our SO(5) model 
consists of spinful fermions and therefore requires a generalization of the fermion parity to 
\begin{equation}
\mathcal P_N (\phi) = \text{sign}\l(
\mathcal E_0^{\sf odd} (\phi) -
\mathcal E_0^{\sf even} (\phi)
\r),
\label{eqn:parity}
\end{equation}

where the ground-state energies are defined as
$\mathcal E_0^{\sf even}(\phi) = \mathcal E_0^\phi(N)$ and $\mathcal E_0^{\sf odd}(\phi) = \frac{1}{2}(\mathcal E_0^\phi (N+2) + \mathcal E_0^\phi (N-2))$ for
fermion number $N$ divisible by four, such that the $N\pm2$-particle states have $N_\uparrow = N_\downarrow$ odd. 
This differs from the SU(2) case (where $\mathcal E_0^{\sf{odd}}$ is the average of $N\pm 1$-particle energies) due to the spin-degeneracy of the $k=0$ mode.
The quantity $\phi=0 \, (2\pi)$ represents periodic (antiperiodic) boundary 
conditions and corresponds to enclosing a flux $\Phi = \frac{\phi \Phi_0}{2\pi}$
in a ring geometry with anomalous flux quantum $\Phi_0$ \cite{ortiz14,ortiz16}. 
In the topologically trivial phase ($g > g_c$),
$\mathcal P_N(\phi)=1$ for both periodic and
antiperiodic boundary conditions (Fig. \ref{fig:derivs}). For $g < g_c$,
a parity switch is observed, with 
$\mathcal P_N(0) = -1$ and $\mathcal P_N(2\pi)=1$.
This can be linked back to occupation of the zero-momentum
state that exists for $\phi=0$: in the
topologically non-trivial phase,
it is energetically 
advantageous to occupy both ($\sigma=\up, \down$) $k=0$ single-particle states instead
of forming a pair. In the pairing-dominated
trivial phase, the $k=0$ states are vacated in
favor of an additional pair at finite momentum.

Unlike what is seen in other SU(2) RG models, there is a
macroscopic degeneracy at the critical point involving multiple 
states from each sector with fixed $N$ and $M=\sum_\k S_\k^z$,
with an accompanying global SO(5) symmetry
generated by operators 
$I^\kappa = \sum_\k I_{\k}^\kappa$ where 
$I_{\k}^\kappa$, $\kappa=1,\dots,10$, is any generator of the SO(5) algebra  in Eq. \eqref{eqn:so5}.
At $g=g_c$, Hamiltonian \eqref{eqn:ham_iom} becomes
the SO(5) Gaudin model
$\sum_{\k,\k'} \eta_\k \eta_{\k'} \vec{\mathcal{T}}_\k \cdot \vec{\mathcal{T}}_{\k'}$, 
and the ground-state solutions to the RG
equations have all pairons $e_\alpha$ equal to zero. Those 
equations then simplify to a single set
for variables $\omega_{\beta}$,
$\sum_{\beta'\neq \beta} Z_{\beta' \beta} = \sum_\k s_\k Z_{\k \beta}$, 
$\beta = 1, \dots, N_\omega$. Each independent solution corresponds to a degenerate eigenstate. 
The entire energy spectrum at this point can be classified according to the 
degenerate SO(5) global irreps constructed from the coupling of the 
$l$ SO(5)$_\k$ irreps $\{\nu_\mathbf{k},s_\mathbf{k}\}$ of each
level. The chain decomposition
$\text{SO(5)}\supset \text{U}_{\bf S}(2)\supset \text{U}_{S^z}(1)$
\cite{rowe12} classifies  the complete set of eigenstates in terms of the
fermion number $N$ and spin content $S$ in each  global irrep. 
The wavefunctions constituting the ground state irrep are
defined in terms of $S=0$ quartet creation operator,
$
Q\p = \sum_{\k,\k'}\l(
T_{1\k}\p T_{-1\k'}\p
+ T_{-1\k}\p T_{1\k'}\p
- T_{0\k}\p T_{0\k'}\p
\r)$,
$S=1$ global pair operators 
$T_{\mu}\p = \sum_\k T_{\mu \k}\p$, and spin lowering operator
$S^- = \sum_\k S_\k\m$. For an even 
number of particles $N \leq L$ ($N>L$ states can be determined by 
particle-hole transformation), these states are
\beq
\ket{N_Q, S, M}= \l( S^-\r)^{S-M}\left( Q^+\right)^{N_Q} \left(T^+_1\right)^S\ket 0
\label{eq:WFeven}.
\eeq
Since $N=4N_Q+2S$, the possible values of spin are $S=N/2,N/2-2,\dots,1 \text{ or  }0$, with 
$S=0$ representing the pure quartet state.
From this, we find the
degeneracy of the even $N$ ground-state 
manifold:
$d_{N,M}^{\sf even}=\left\lfloor\frac{\min(N,2L-N)-2|M|}{4}\right\rfloor+1$,
where $\lfloor x\rfloor$
is the largest integer less than or equal to
$x$. The energy of these states is $\mathcal E^{\sf{even}} =-3 \frac{g_c}{L} \sum_\k \eta_\k^2$.

The $N+1$ ($N$ even) particle ground-state irrep has $N$ particles in a wavefunction of the form \eqref{eq:WFeven} with  spin $S_e$  and  one unpaired particle in the lowest momentum level $k_m$, giving total spin $S=S_e\pm 1/2$ with
possible values $S=N/2,N/2-1,\dots, 1/2$.
The number of particles is then $N=4 N_Q+2S_e+1$.
From the available spins and the additional two-fold degeneracy  
arising from  the two momenta ($\pm \mathbf{k}_m$) of  the unpaired particle, 
the degeneracy of the  odd-sector ground-state subspace
is:  
$
d^{\sf odd}_{N,M}=\frac{\min(N,2L-N)-2|M|}{2}+1$.
The energy of these states
$\mathcal E^{\sf{odd}} = \mathcal E^{\sf{even}}+ \eta_{k_m}\l(
1+\frac{g_c}{2L}\eta_{k_m}\r)$ simplifies to the even-$N$ energy
plus the kinetic energy of the unpaired fermion in the
thermodynamic limit.

The presence of quartets in a Hamiltonian such as \eqref{eqn:ham_iom} deserves 
 mention. We are only aware of 
the significance of quartet correlations in atomic nuclei \cite{Physicality.87.192501, PhysRevC.85.061303} and in exotic phases of cold spin-3/2 fermionic atoms \cite{wu03, wu05}. It is
important, then, to establish the interactions that take our system away from its $g=g_c$ critical point and stabilize 
a quartet, as opposed to a paired, ground state. To this end, we compare the quartet, $\Delta_4(N) = 
\l (\mathcal E_0(N+2) + \mathcal E_0(N-2) - 2\mathcal E_0(N)\r )/2$ (see for example \cite{PhysRevC.90.024322}), and paired, $\Delta_2(N) = \l(\mathcal E_0(N+1) +
\mathcal E_0(N-1) - 2\mathcal E_0(N)\r)/2$, gaps in our Hamiltonian \eqref{eqn:ham_iom} as a function of $g$. If
$\Delta_4(N) \sim \Delta_2(N)$, we say that there are significant quartet correlations in the ground state for that value of $g$. 
Our analysis indicates that 
quartet correlations become more relevant in the repulsive sector and, in the attractive sector, for pairing-only (non-integrable) interactions \cite{SM}.

{\it SO(5) magnetic superfluid} -- 
An interesting physical mechanism emerges when our SO(5) system  \eqref{eqn:ham_iom} is subject to an external magnetic field $h$ as in Eq. \eqref{eqn:hamiltonian}. At low temperatures and 
pressures superfluid $^3$He, known to have both $p$-wave pairing and ferromagnetic
interactions \cite{leggett75}, displays transitions between non-magnetic (B) and 
magnetic (A) superfluid phases as function of an applied magnetic field. 
The A and B phases of superfluid $^3$He are associated
with the mean-field wavefunctions proposed by Anderson, Brinkman, and Morel
(ABM) and Balian and Werthamer (BW), respectively \cite{leggett75}.
The BW state is a simple generalization of BCS principles to spin-triplet
(rather than spin-singlet) pairs, and in the SO(5) language 
is a superposition of $T_{-1}\p$, $T_0\p$, and $T_1\p$ operators
acting on the vacuum. The ABM state is structured similarly, but
allows only like-spin fermion pairs,
ruling out the channel generated by $T_0\p$.
Experimentally, it is known
that in absence of a magnetic field, 
the B phase is the only possible superfluid at 
zero temperature. With the addition of a magnetic 
field, both phases become accessible at zero temperature
along with the spin-polarized superfluid A1 phase \cite{lee97}.

\begin{figure}
\includegraphics[scale=1]{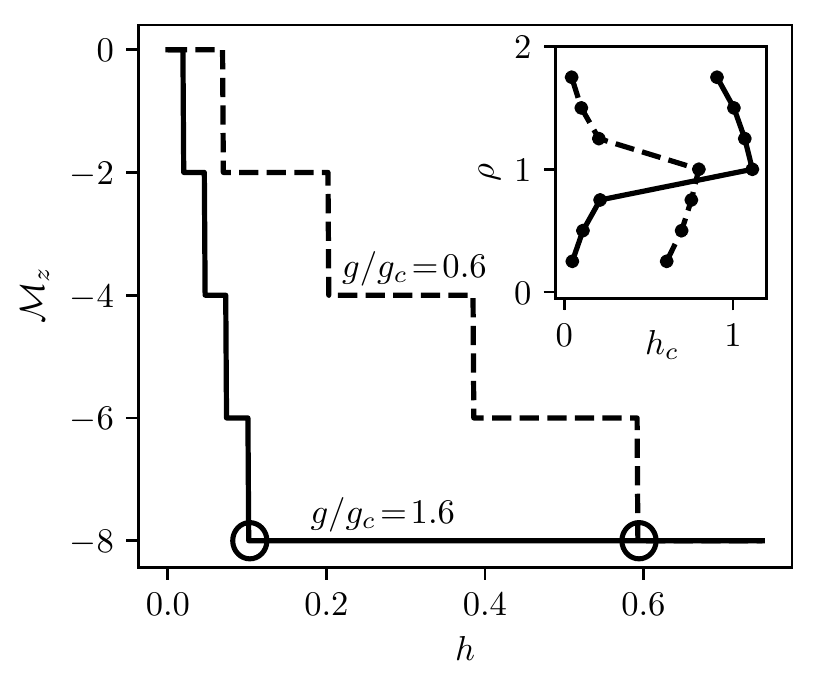}
\caption{
Ground state magnetization
${\cal M}_z$ as a function
of $h$ at $g/g_c=0.6$ (dashed line) and 
1.6 (solid line) for $L = 32$ and 
$N = 16$ ($\rho=1/2$). The circled points are at 
$h=h_c$, the smallest field that fully polarizes the system. Inset: Density $\rho$ as a function of the thermodynamic extrapolation of $h_c$, indicating a phase boundary between
fully and semi-polarized magnetic superfluids.
}
\label{fig:h_c}
\end{figure}

Interestingly, 
our model demonstrates a series of first-order 
magnetically-driven transitions 
between different spin-triplet superfluids with
no pair-breaking, which may provide insight into magnetic superfluidity.
To determine the ground state energy of the Hamiltonian \eqref{eqn:ham_iom}, it suffices to 
find $\mathcal E_0(N,M)$, the lowest energy of the
$h=0$ Hamiltonian for each possible value of $S^z$ (a conserved quantity)
and determine which value of $M$ gives
the lowest total energy
$
\mathcal E_0(h) = \min_{M}\l(\mathcal E_0(N,M) - h M\r)$.
This process simplifies for $2g  > g_c$,
since above this coupling
we find the ground state has no unpaired fermions for any value of $h$.
From this formula, it is clear that the magnetization 
$\mathcal M_z= \partial \mathcal E_0/\partial h$ is equal to $M$.
Rather than breaking pairs, the magnetic field changes the balance of $-1$, 0, and
1 pairs. 
This manifests as a 
series of first-order phase transitions as $\mathcal M_z$
jumps between integers with the same parity as $N_e$,
as illustrated in Fig \ref{fig:h_c}.
The minimum value of $h$ for nonzero magnetization goes to zero in the
thermodynamic limit,
while the final transition occurs at a value, $h=h_c$,
that remains finite in that limit.
A similar type of mechanism may be at play in superfluid $^3$He
leading to the emergence of the A1 phase.
Crucially, the $h\neq 0$
ground states of our model are merely the lowest-energy solutions to the 
$h=0$ problem at the same coupling $g$ in a sector with $N_\uparrow\neq N_\downarrow$, and so shares 
topological and superfluid properties with the $h=0$ ground state at the same coupling $g$.

This non pair-breaking mechanism, already encoded in the {\it exact} 
solution, can be modeled at the {\it mean-field} level by introducing the 
SO(5) generalized coherent state \cite{Perelomov_1986}
\beq
\ket \Psi = e^{\sum_k z_k \l(x_1 \, T_{0k}\p + x_2 T_{1k}\p
+ x_3 T_{-1k}\p \r)} \ket 0 ,
\eeq
where $x_{1}^2+x_{2}^2+x_{3}^2=1$ and $\{z_k\}$ are variational parameters \cite{SM}.
As $|x_{1,2,3}|$ goes from 0 to 1, this state goes
from having only $M=0$ pairs through
a state with a mixture of all three pairing channels, similar
to the BW state, to a state with only like-spin
pairs, similar to the ABM wavefunction.

{\it Metallic phases} --- 
For repulsive couplings ($g < 0$), the
ground state has a momentum distribution with
a discontinuity at the Fermi momentum $k_F$ (Fig. \ref{fig:phases}), suggesting
a ground state almost identical to a noninteracting Fermi gas
$
\ket{\Psi_{\sf{nonint}}} = \frac{1}{\sqrt{N!}}\prod_{k=-k_F}^{k_F}c_{k\up}\d
c_{k\down}\d\ket 0$ with energy 
$\mathcal E_{\sf{nonint}}(N)=\bra{\Psi_{\sf{nonint}}} H \ket{\Psi_{\sf{nonint}}}$.
This discontinuity persists even for strongly repulsive couplings,
unlike what is usually observed in an interacting Fermi liquid \cite{ortiz05,Baym_1991}.
In the thermodynamic limit, 
the ground state energy density 
$e_0$ converges to $e_{\sf{nonint}}=\frac{\pi}{4}\rho^2 - \frac{g\pi^2}{64}\rho^4
+ \mathcal O(1/L)$ \cite{SM}.

One may wonder whether \eqref{eqn:ham_iom} has a flat-band limit in the strong coupling
($g\rightarrow -\infty$) limit. The SU(2) RG model shares similar metallic properties with
the SO(5) model for low couplings,
but the flat-band Hamiltonian $\lim_{g\rightarrow -\infty}\frac{1}{g}H$ has
an exponentially degenerate ground-state manifold \cite{SM}. 
This limiting case has been studied, for instance, in fractional quantum 
Hall liquids \cite{ortiz13}, and its importance lies in the
non-Fermi liquid behavior that manifests due to a high density of states near
the ground state. 
Due to the presence of effective single-particle
terms in the interaction, the  SO(5) model in 
\eqref{eqn:ham_iom} does not exhibit high degeneracy in this
limit. Instead, a level crossing occurs at a nonuniversal
coupling where the ground state gains a nonzero seniority independent of system size \cite{SM}.
By removing all single-particle terms, one arrives
at a special case of the Hamiltonian 
\eqref{eqn:hamiltonian} with an exponentially degenerate ground state in
the flat band (pure interaction) limit.
A detailed discussion of this behavior is beyond the scope of 
this paper \cite{SM}.

{\it Concluding Remarks} --- 
We have presented an exactly-solvable model displaying SO(5) topological superfluidity.
Its relevance lies in providing a new non-pair-breaking mechanism for magnetic superfluids, 
of relevance for liquid $^3$He or other exotic spin-triplet $p$-wave superfluids. 
At a critical coupling separating trivial and non-trivial topological superfluids, 
the model reduces to an (global) SO(5) Gaudin Hamiltonian. These phases show 
quartet correlations that become more prevalent as magnetic and density interactions 
are quenched. The repulsive phases of the model are also of interest, in particular, 
in relation to non-Fermi liquid behavior; they deserves further study.
Finally, we would like to make connection to a seemingly unrelated phenomenon. 
The positive semi-definite (frustration-free) Haldane-Rezayi Hamiltonian \cite{haldane88,Weerasinghe_2014} 
$H=\sum_{0<j<L} H_j$,  
$H_j=\sum_{k,k'}\eta_k \eta_{k'}
\vec{T}^+_k\!\cdot\! \vec{T}^-_{k'}$ with
$T^+_{0,k}=(c_{j+k\up}\d c_{j-k\down}\d + c_{j+k\down}\d c_{j-k\up}\d)/\sqrt{2}$, 
$T^+_{1,k}=c_{j+k\up}\d c_{j-k\up}\d$, 
$T^+_{-1,k}=c_{j+k\down}\d c_{j-k\down}\d$,
defined in a cylinder ($k,k'\in [-j,j]$ are angular momenta indexes),
stabilizes a gapless zero mode at filling fraction $\nu=1/2$ representing a non-Abelian fractional 
quantum Hall trial state \cite{Seidel_2011}.
We have shown that (positive semi-definite) Hamiltonian $H_j$ 
is an element of SO(5) but, as a 
corollary of this work, it is not integrable \`a la RG. 
As it is a (repulsive) pairing-only Hamiltonian,
it is expected that quartet correlations become relevant. 
Interestingly, each $H_j$ has a macroscopically degenerate zero-energy subspace and the 
intersection of their kernels result into the Haldane-Rezayi state. 

\begin{acknowledgments}
We acknowledge illuminating discussions with Henri Godfrin and Erkki Thuneberg on
the nature of phase transitions in Superfluid $^3$He subject to strong magnetic fields, 
and Grigory Volovik for pointing out Ref. \cite{Hasegawa1979}.
The open-source Python package \textsc{QuSpin}\cite{weinberg17} was used for exact diagonalization.

S.L.-H. acknowledges financial support from the Mexican CONACyT project CB2015-01/255702.
J.D. is supported by the Spanish Ministerio de Ciencia e Innovaci\'on,
and the European regional development fund (FEDER) under Project No.
PGC2018-094180-B-I00, S.L.-H. and J.D. acknowledges financial support from the Spanish 
collaboration Grant I-COOP2017 Ref:COOPB20289. G.O. and W.H. acknowledge
support from the US Department of Energy grant  DE-SC0020343.
\end{acknowledgments}

\end{document}

\title{Supplemental Material: \\
Integrable Model of Topological SO(5) Superfluidity}

\author{Will J. Holdhusen}
\affiliation{Department of Physics, Indiana University, Bloomington, Indiana 47405, USA}

\author{Sergio Lerma-Hern\'andez}
\affiliation{Facultad de F\'{i}sica, Universidad Veracruzana, Circuito Aguirre Beltr\'an s/n, Xalapa, Veracruz 91000, Mexico}

\author{Jorge Dukelsky}
\affiliation{Instituto de Estructura de la Materia, CSIC, Serrano 123, 28006 Madrid, Spain}

\author{Gerardo Ortiz}
\affiliation{Department of Physics, Indiana University, Bloomington, Indiana 47405, USA}

\maketitle

\tableofcontents

\section{The SO(5) algebra: commutation relations and fermionic representation}
\label{app:so5}
The generators of SO(5) satisfy the commutation 
relations
\begin{align}
[S_\k\z,  \tv_{\k'}\p]
&= \delta_{\k,\k'}\bx  -T_{-1 \k}\p\\ 0 \\ T_{1 \k}\p \ex \\
[S_\k\z, S_{\k'}^\pm]&=\pm \delta_{\k,\k'}S_\k^{\pm}\\
[S_\k\p,  \tv_{\k'}\p]
&=\sqrt 2 \delta_{\k,{\k'}}
\bx T_{0\k}\p \\ T_{1\k}\p \\ 0\ex\\
[S_\k\p,S_{\k'}\m]
&=2\delta_{\k,{\k'}} S_\k\z\\
[S_\k\m, \tv_{\k'}\p]
&=\sqrt 2 \delta_{\k,{\k'}}
\bx 0\\ T_{-1\k}\p\\ T_{0\k}\p \ex\\
[N_{\k}\n, \tv_{{\k'}}\p]
&=2\delta_{\k,{\k'}}\tv_\k\p\\
[T_{-1\k}\m,  \tv_{{\k'}}\p]
&=\f{1}{\sqrt 2}\delta_{\k,{\k'}}\bx
\sqrt 2(1 + S_\k^z - \f{1}{2}N_\k\n)\\
-S_\k\p\\
0
\ex\\
[T_{0\k}\m, \tv_{{\k'}}\p]
&=\f{1}{\sqrt 2}\delta_{\k,{\k'}}
\bx
-S_\k\m\\
\sqrt 2 \l(1-\f{1}{2}N_\k\n\r)\\
-S_\k\p
\ex\\
[T_{1\k}\m, \tv_{{\k'}}\p]
&=\f{1}{\sqrt 2}\delta_{\k,{\k'}}
\bx
0 \\
-S_\k\m\\
\sqrt 2(1 - S_\k^z  - \f{1}{2}N_\k\n)
\ex
\end{align}
with all other commutators between these operators
equal to zero. 
Here, we use notation
$\vec T_\k\p = \l(T_{-1\k}\p, T_{0\k}\p, 
T_{1\k}\p\r)$ for the spherical components of the 
vector operator.

These generators form four (non-commuting) SU(2) subalgebras
on each level:
one from the spin operators $\{S_\k\p, S_\k\m, S_\k\z\}$, 
and one from each pair creation operator:
\beq
\l\{T_{\mu \k}\p, T_{\mu \k}\m, \f{1}{2}N_\k\n-1 + \mu S_\k\z\r\}, 
\quad\mu \in \{-1, 0, 1\}.
\eeq
The SO(5) algebra has Casimir operators
\beq
\mathcal{T}_\k^2 = \t_\k\p \cdot \t_\k\m
+ \t_\k\m \cdot \t_\k\p
+ S_\k^2
+ \l(\f{N_\k}{2}-1\r)^2.\label{eqn:casimir}
\eeq
where 
$S_\k^2= \frac{1}{2}\l(S_\k\p S_\k\m + S_\k\m S_\k\p\r) + S_\k\z S_\k\z$.
These commute with all elements of the algebra, and have
eigenvalues for each level 
determined by seniority and reduced spin quantum numbers $\nu_\k$ and $s_\k$:
\beq
\braket{\mathcal T_\k^2} =
s_\k\n(s_\k\n+1) + \l(\frac{\nu_\k}{2}-1\r)^2 - 3
\l(\f{\nu_\k}{2}-1\r)
\equiv t_\k^2. \label{eqn:casimir_val}
\eeq
The generators of a local (acting on modes at $\{+\k, -\k\}$) SO(5) algebra can be written in a fermionic ($c_{\k\sigma}\d$) representation as
\begin{subequations}
\begin{eqnarray}
T_{\pm 1 \k}\p &=& c_{\k\sigma_\pm}\d c_{-\k\sigma_\pm}\d= (T_{\pm 1 \k}\m)\d,\\
T_{0\k}\p &=&\f{1}{\sqrt 2}\l(
c_{\k\up}\d c_{-\k\down}\d + c_{\k\down}\d c_{-\k\up}\d\r)= (T_{0 \k}\m)\d,\\
N_{\k}\n &=&n_{\k\up}\n + n_{-\k\up}\n + n_{\k\down}\n
+ n_{-\k\down}\n,\\
S_\k\z &=&\f{1}{2}\l(
n_{\k\up}\n + n_{-\k\up}\n - n_{\k\down}\n - n_{-\k\down}\n
\r), \\
S_\k\p &=&
c_{\k\up}\d c_{\k\down}\n + c_{-\k\up}\d c_{-\k\down}\n
= (S_\k\m)\d
\label{eqn:so5rep}
\end{eqnarray}
\end{subequations}
with $\sigma_+=\up, \sigma_{-}=\down$.

\section{The SO(5) Richardson-Gaudin Hamiltonian}
\label{app:hamiltonians}

\begin{figure}
\centering
\includegraphics{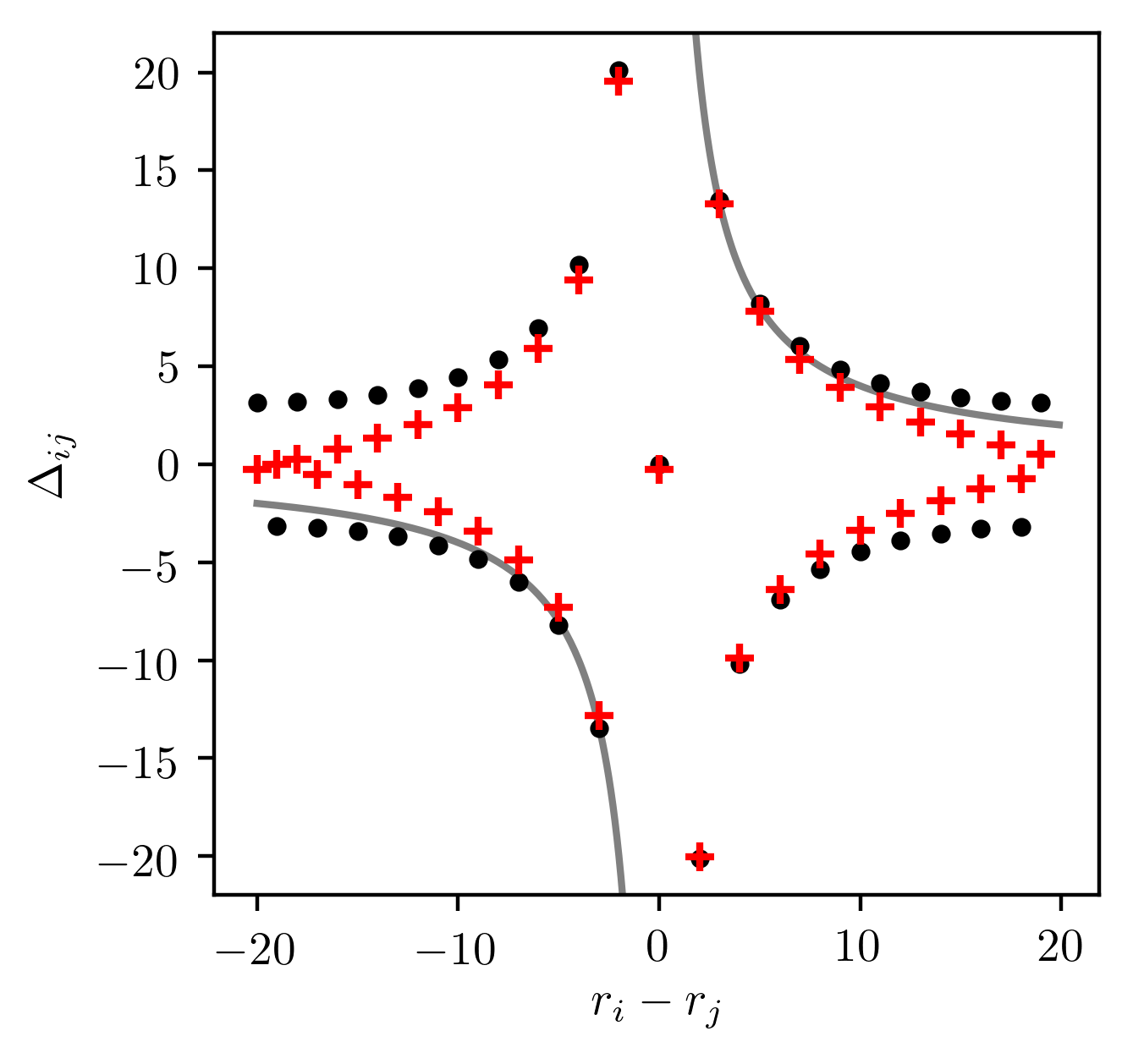}
\caption{Spatial pairing coefficients $\Delta_{ij}$
(Eq. \eqref{eq:coeffs} as a function of the displacement $r_i-r_j$ with $r_i = i$ (lattice constant set to unity) for periodic (crosses) and antiperiodic (dots) boundary conditions in a system of size $L=40$. The gray line is $L/(r_i - r_j)$, showing the $1/r$ decay of $|\Delta_{ij}|$.}
\label{fig:coeffs}
\end{figure}

The exactly-solvable Hamiltonian
\beq
H = 
\sum_{\k} \eta_\k N_\k
- \frac{g}{L} \sum_{\k\k'}\eta_\k \eta_{\k'}
\l(
\vec T_\k\p \cdot \vec T_{\k'}\m + \vec T_\k\m \cdot \vec T_{\k'}\p
+ \vec S_\k\cdot \vec S_{\k'}
+ \frac{1}{4} N_\k\n N_{\k'}\n
\r) \label{eqn:hamiltonian}
\eeq
can be expressed
as a linear combination of the RG integrals
of motion
\beq
R_\k = \frac{1}{2}N_\k-1
+ q \sum_{\k'\neq \k}Z_{\k\k'}\vec{\mathcal T_\k} \cdot \vec{\mathcal T_{\k'}}
\eeq
with
$\vec{\mathcal T_\k\n} \cdot \vec{\mathcal T_{\k'}\n}
=\vec T_\k\p \cdot \vec T_{\k'}\m
+ \vec T_\k\m \cdot \vec T_{\k'}\p
+ \vec S_\k\n\cdot \vec S_{\k'}\n +
\l(N_\k\n/2-1\r)\l(N_{\k'}\n/2-1\r)$.

Summing the integrals of motion with coefficients $\eta_\k$
and using the antisymmetry of the function 
$Z_{\k\k'}=\frac{\eta_\k \eta_{\k'}}{\eta_\k - \eta_{\k'}}$, we arrive at
\begin{align*}
\sum_\k \eta_\k R_\k 
=&
\frac{1}{2}\sum_\k \eta_\k \l(N_\k-2\r)
+ \frac{q}{2}\sum_{\k,\k'\neq \k}
\l(\eta_\k - \eta_{\k'}\r)Z_{\k\k'}
\vec{\mathcal T_\k} \cdot \vec{\mathcal T_{\k'}}
=\frac{1}{2}\sum_\k \eta_\k \l(N_\k-2\r)
+\frac{q}{2}\sum_{\k, \k' \neq \k}\eta_\k\eta_{\k'}
\vec{\mathcal T_\k} \cdot \vec{\mathcal T_{\k'}}.
\end{align*}
Now, we add in the diagonal elements of the double sum,
which are the SO(5) Casimirs \eqref{eqn:casimir} whose
eigenvalues are the quantum numbers $t_\k^2$
\eqref{eqn:casimir_val},
and arrive at
\begin{align}
\begin{split}
\sum_\k \eta_\k R_\k=&
\frac{1}{2}\l(1 - q\sum_{\k'} \eta_{\k'}\r)
\sum_\k \eta_\k N_\k
+\f{q}{2}\sum_{\k\k'}\eta_\k\eta_{\k'}\l(
\vec T_\k\p \cdot \vec T_{\k'}\m + \vec T_{\k}\m \cdot \vec T_{\k'}\p
+ \vec S_\k \cdot \vec S_{\k'} + \frac{1}{4}N_\k N_{\k'}
\r) + \Gamma(q).
\end{split}
\end{align}
where $\Gamma(q)=- \sum_\k\eta_\k\l(
1-\f{q}{2} \sum_{\k'}\eta_{\k'}
+ \f{q}{2}\eta_\k t_\k^2\r)$.
Subtracting this constant, rescaling by a factor of 
$2/(1-q\sum_{\k'}\eta_{\k'})$,
and defining the physical coupling 
$g/L  \equiv -q/(1-q\sum_{\k'}\eta_{\k'})$
gives the Hamiltonian \eqref{eqn:hamiltonian}.

In one spatial dimension, we can take
the Fourier transform of the interaction coefficients
$\delta_{kk'}=\eta_k\eta_{k'}$ to understand their behavior in coordinate
space:
\beq
\Delta_{ij} = \frac{1}{L}\sum_{k,k'} \eta_k \eta_{k'} e^{ikr_i} e^{ik'r_j}.
\label{eq:coeffs}
\eeq
In Fig. \ref{fig:coeffs}, we plot the behavior of these coefficients for $\eta_k = k$ with periodic and antiperiodic boundary conditions. As is seen in the plot, these behave
as $\Delta_{ij} \propto (-1)^{i-j}/(r_i-r_j)$.

\section{SO(5) Richardson-Gaudin wavefunctions}
It is difficult to intuitively understand the Richardson-Gaudin (RG)
wavefunctions from their formal description.
To illustrate the form of these wavefunctions, it is useful to consider
small systems. As a simplification, we will restrict ourselves
to the zero-seniority case, which removes the $\S_\beta^+$ operator.

First, consider cases in which $N_e = N_\omega$,
corresponding to the ground state sector when there is no
external magnetic field.

For $N_e = N_\omega = 1$, the eigenstates are simply
\begin{align*}
\ket{\Psi}
=Z(e_1\n,\omega_1\n)\sum_\k Z(\eta_\k\n, e_\k\n) T_{0\k}\p \ket 0.
\end{align*}
For $N_e = N_\omega = 2$,
the wavefunction is already much more complicated:
\beq
\ket \Psi
 =\sum_{\k_1,\k_2}\l(
C^{1-1}_{\k_1\k_2}T_{1\k_1}^+ T_{-1\k_2}^+
+ C^{00}_{\k_1\k_2}T_{0\k_1}^+ T_{0\k_2}^+
\r)\ket 0 \label{eqn:NeNw2}
\eeq
where the coefficients are
\begin{subequations}
\begin{align}
C^{1-1}_{\k_1 \k_2} =& Z(e_1, \omega_1)Z(e_1, \omega_2)
Z(\eta_{\k_1}, e_1)Z(\eta_{\k_2}e_2) 
+ Z(e_2,\omega_1) Z(e_2,\omega_2)
Z(\eta_{\k_1}, e_2) Z(\eta_{\k_2}, e_1),
\\
C^{00}_{\k_1 \k_2} =&Z(e_1, \omega_1) Z(e_2, \omega_2)
Z(\eta_{\k_1}, e_1)Z(\eta_{\k_2}, e_2)
+Z(e_2, \omega_1)Z(e_1, \omega_2)
Z(\eta_{\k_1}, e_1)Z(\eta_{\k_2}e_2).
\end{align}
\end{subequations}
It can be shown that these coefficients obey the relation
$\sum_{\k_1 \k_2} C^{1-1}_{\k_1 \k_2} = -2 \sum_{\k_1 \k_2} C_{\k_1\k_2}^{00}$.

Moving one more size up, the
$N_e=  N_\omega = 3$ wavefunction can be written
\beq
\ket{\Psi} = \sum_{\k_1 \k_2 \k_3}\l(
C_{\k_1 \k_2 \k_3}^{10-1}T_{1\k_1}\p T_{0\k_2}\p T_{-1\k_3}\p
+C_{\k_1 \k_2 \k_3}^{000}T_{0\k_1}\p T_{0\k_2}\p T_{0\k_3}\p\r)
\eeq
To express these coefficients concisely, we write them
as permanents and sum over the symmetric group $S_3$ to
consider permutations of $\{\omega_1, \omega_2, \omega_3\}$:
\begin{align}
C_{\k_1 \k_2 \k_3}^{000}
=& \text{perm}
\begin{pmatrix}
Z(e_1,\omega_1) Z(\eta_{\k_1}, e_1) 
& Z(e_2,\omega_1) Z(\eta_{\k_1}, e_2) 
& Z(e_3,\omega_1) Z(\eta_{\k_1}, e_3) \\
Z(e_1,\omega_2) Z(\eta_{\k_2}, e_1) 
& Z(e_2,\omega_2) Z(\eta_{\k_2}, e_2) 
& Z(e_3,\omega_2) Z(\eta_{\k_2}, e_3) \\
Z(e_1,\omega_3) Z(\eta_{\k_3}, e_1) 
& Z(e_2,\omega_3) Z(\eta_{\k_3}, e_2) 
& Z(e_3,\omega_3) Z(\eta_{\k_3}, e_3) \\
\end{pmatrix}\\
C_{\k_1 \k_2 \k_3}^{10-1}
=& \!\!\sum_{\{a, b, c\}\in S_3}\!\!
\text{perm}
\begin{pmatrix}
Z(e_1,\omega_a) Z(e_1,\omega_b) Z(\eta_{\k_1}, e_1)
& Z(e_2,\omega_a) Z(e_2,\omega_b) Z(\eta_{\k_1}, e_2)
& Z(e_3,\omega_a) Z(e_3,\omega_b) Z(\eta_{\k_1}, e_3)\\
Z(e_1, \omega_c) Z(\eta_{\k_2}, e_1)
& Z(e_2, \omega_c) Z(\eta_{\k_2}, e_2) 
& Z(e_3, \omega_c) Z(\eta_{\k_2}, e_3)\\
Z(\eta_{\k_3}, e_1) & Z(\eta_{\k_3}, e_2) & Z(\eta_{\k_3}, e_3)
\end{pmatrix}
\end{align}
This permanent structure is also present in the previously
considered eigenstates, and in fact the creation operators
for these states may also be expressed as permanents.

Unlike singlet-pairing states, eigenstates of our Hamiltonian
can be fully or partially spin polarized without breaking
pairs due to the different $S^z$ pairing channels.
The $N_e=2$ states are fairly simple:
Starting with the trivial case $N_e = 2$, $N_\omega = 0$,
\beq
\ket{\Psi} = \sum_{\k_1 \k_2}Z(\eta_{\k_1}\n, e_1\n) Z(\eta_{\k_2}\n, e_2\n) T_{-1\k_1}\p T_{-1\k_2}\p.
\eeq
Now with $N_e=2$ and $N_\omega = 1$, we get
\beq
\ket{\Psi} 
= \sum_{\k_1\k_2}C_{\k_1 \k_2}^{0-1}T_{0\k_1}\p T_{-1\k_2}\p
\eeq
The spin-up polarized wavefunctions will
have equivalent forms with $-1$ and $1$ subscripts
interchanged.

As a final example, consider the case $N_e = 3$, 
$N_\omega = 2$. Here, the wavefunction is
\beq
\ket{\Psi}=
\sum_{\k_1 \k_2 \k_3}\l(C_{\k_1 \k_2 \k_3}^{1-1-1}T_{1\k_1}\p T_{-1\k_2}\p T_{-1\k_3}\p
+ C_{\k_1\k_2\k_3}^{00-1}
T_{-1\k_1}\p T_{0\k_2}\p T_{0\k_3}\p
\r)
\eeq
with coefficients
\begin{align}
C_{\k_1\k_2\k_3}^{1-1-1} = &\frac{1}{2} \text{perm}\begin{pmatrix}
Z(e_1,\omega_1) Z(e_1,\omega_2) Z(\eta_{\k_1}, e_1) 
& Z(e_2,\omega_1) Z(e_2,\omega_2)Z(\eta_{\k_1}, e_1) 
& Z(e_3, \omega_1) Z(e_2,\omega_1)Z(\eta_{\k_1}, e_3)\\
Z(\eta_{\k_2}, e_1) & Z(\eta_{\k_2}, e_2) & Z(\eta_{\k_2}, e_3)\\
Z(\eta_{\k_3}, e_1) & Z(\eta_{\k_3}, e_2) & Z(\eta_{\k_3}, e_3)
\end{pmatrix}\\
C_{\k_1\k_2\k_3}^{00-1}
= &\text{perm} \begin{pmatrix}
Z(e_1, \omega_1) Z(\eta_{\k_1}, e_1) 
& Z(e_2, \omega_1) Z(\eta_{\k_1}, e_2)
& Z(e_3, \omega_1) Z(\eta_{\k_1}, e_3)\\
Z(e_1, \omega_2) Z(\eta_{\k_2}, e_1) 
& Z(e_2,\omega_2) Z(\eta_{\k_2}, e_2) 
& Z(e_3,\omega_2)Z(\eta_{\k_2}, e_3) \\
Z(\eta_{\k_3}, e_1) & Z(\eta_{\k_3}, e_2) & Z(\eta_{\k_3}, e_3)
\end{pmatrix}.
\end{align}
Similarly to the $N_e = N_\omega = 2$ case,
these coefficients satisfy a relationship
$\sum_{\k_1\k_2\k_3}C_{\k_1\k_2\k_3}^{1-1-1} =
-2\sum_{\k_1\k_2\k_3}C_{\k_1\k_2\k_3}^{00-1}$.

\section{SO(5) mean-field solutions}

\label{app:mf} 
We start with the SO(5) coherent state
\begin{equation}
\ket\Psi =e^{\sqrt{2}\sum_{k}z_{k}\Gamma _{k}^{\dagger }}\ket0
\label{eqn:mf_wf}
\end{equation}%
where $\Gamma _{k}^{\dagger }=\sqrt{\left( 1-x\right) }T_{0k}^{+}+\sqrt{(x/2)%
}\left( T_{1k}^{+}-T_{-1k}^{+}\right) $. Variational parameters $\{z_{k}\}$
and $x$ can be chosen to minimize the energy of this wavefunction with the
Hamiltonian \eqref{eqn:hamiltonian}.

The $\{z_{k}\}$ variables related to the $u_{k}$ and $v_{k}$ factors used in
BCS superconductivity as $z_{k}=v_{k}/u_{k}$ with restriction $%
|u_{k}|^{2}+|v_{k}|^{2}=1$. As $x$ goes from 0 to 1, the wavefunction goes
from having only $S^{z}=0$ pairs through a state with a mixture of all three
pairing channels similar to the BW state to a state with only $S^z=\pm 1$
pairs, similar to the ABM wavefunction. For $x\neq 0$, the state 
\eqref{eqn:mf_wf} is no longer an eigenstate of $S^{z}$ but preserves the
zero magnetization $\left\langle S^{z}\right\rangle =0$.

To approximate the ground state, we minimize the expectation value of $%
H^{\prime }=H-\lambda \sum_{k}N_{k}$, where $H$ is the integrable Hamiltonian of 
Eq. (9) of the main paper 
and $\lambda $ is a Lagrange
multiplier fixing the particle number $\bra{\Psi}\sum_{k}N_{k}\ket\Psi =N$.
Simple calculations show
\begin{equation*}
\braket{\Psi|H'|\Psi}=4\sum_{k}\left( \eta _{k}-\lambda \right) \frac{%
z_{k}^{2}}{\left( 1+z_{k}^{2}\right) }-4G\sum_{kk^{\prime }}\eta _{k}\eta
_{k^{\prime }}z_{k}z_{k^{\prime }}\frac{1+z_{k}z_{k^{\prime }}}{\left(
1+z_{k}^{2}\right) \left( 1+z_{k^{\prime }}^{2}\right) }.
\end{equation*}%
Note that the energy is independent of $x$, meaning that the mean-field state is 
a degenerate mixture of the three spin components with $\langle S^z \rangle=0$. 

By taking the derivatives with respect to the variational parameters $z_{k}$
and setting them to zero, we find the extremum conditions that give the set
of mean-field equations (for all $k>0$) plus constraint equation
\begin{subequations}
\begin{align}
0 =&\left( \eta _{k}-\lambda \right) z_{k}-G\eta _{k}\sum_{k^{\prime }}\eta
_{k^{\prime }}z_{k^{\prime }}\frac{\left( 1-z_{k}^{2}+2z_{k}z_{k^{\prime
}}\right) }{\left( 1+z_{k^{\prime }}^{2}\right) }=0  \label{gap0}\\
N =& \sum_k \frac{z_k^2}{1+z_k^2} \label{number0}.
\end{align}
\end{subequations}

These are the set of mean-field equations that define the parameters $%
\left\{ z_{k}\right\} $ and $\lambda $. However, it is more clear and
convenient to express them in terms of the more familiar $\{u_{k},v_{k}\}$
BCS factors. Equations \eqref{gap0} and \eqref{number0} transform to
\begin{subequations}
\begin{align}
0=&\left( \eta _{k}-\lambda \right) v_{k}u_{k}-G\eta _{k}\left(
u_{k}^{2}-v_{k}^{2}\right) \sum_{k^{\prime }}\eta _{k^{\prime }}u_{k^{\prime
}}v_{k^{\prime }}-2G\eta _{k}u_{k}v_{k}\sum_{k^{\prime }}\eta _{k^{\prime
}}v_{k^{\prime }}^{2} \label{gap1}\\
N =& 4\sum_k v_k^2 \label{number1}.
\end{align}
\end{subequations}
We now define the gap $\Delta $, the Fock term $\Gamma $, and the
single-particle Hartree-Fock (HF) energies $\varepsilon _{k}$ as
\begin{equation*}
\Delta =2G\sum_{k}\eta _{k}u_{k}v_{k}, \quad \Gamma =-G\sum_{k}\eta
_{k}v_{k}^{2}, \quad \varepsilon _{k}=\eta _{k}\left( 1+2\Gamma \right)
-\lambda .
\end{equation*}
Inserting in (\ref{gap1}) and solving for $\left\{ u_{k},v_{k}\right\}$,
\begin{equation}
u_{k}^{2}=\frac{1}{2}\left[ 1+\frac{\varepsilon _{k}}{\sqrt{\varepsilon
_{k}^{2}+\eta _{k}^{2}\Delta ^{2}}}\right] \text{, \ }v_{k}^{2}=\frac{1}{2}%
\left[ 1-\frac{\varepsilon _{k}}{\sqrt{\varepsilon _{k}^{2}+\eta
_{k}^{2}\Delta ^{2}}}\right]
\end{equation}
The mean-field equations for the three unknowns $\Delta \,$, $\Gamma $ and $%
\lambda $ are then
\begin{subequations}
\begin{align}
1=&G\sum_{k}\frac{\eta _{k}^{2}}{\sqrt{\left[ \eta _{k}\left( 1+2\Gamma
\right) -\lambda \right] ^{2}+\eta _{k}^{2}\Delta ^{2}}},  \label{Gap}\\
N=&L-2\sum_{k}\frac{\eta _{k}\left( 1+2\Gamma \right) -\lambda }{\sqrt{\left[
\eta _{k}\left( 1+2\Gamma \right) -\lambda \right] ^{2}+\eta _{k}^{2}\Delta
^{2}}},  \label{Number}\\
\Gamma =&-\frac{G}{2}\sum_{k}\eta _{k}\left[ 1-\frac{\eta _{k}\left( 1+2\Gamma
\right) -\lambda }{\sqrt{\left[ \eta _{k}\left( 1+2\Gamma \right) -\lambda %
\right] ^{2}+\eta _{k}^{2}\Delta ^{2}}}\right] .  \label{Fock}
\end{align}
\end{subequations}
Once the system of equations is solved, the ground state energy can be
evaluated in terms of these parameters as
\begin{equation*}
\mathcal E=-\frac{1}{G}\l(\Delta^2 + 4\Gamma(1+\Gamma)\r)
\end{equation*}
In the thermodynamic limit, this becomes a set of integral equations
\begin{subequations}
\begin{align}
1=&\frac{g}{2\pi}\int_0^\pi \frac{\eta_k^2}{\sqrt{[\eta_k(1+2\Gamma)-\lambda]^2 + \eta_k^2 \Delta^2}}dk\\
\rho = &1-\frac{1}{2\pi}\int_0^\pi 
\frac{\eta _{k}\left( 1+2\Gamma \right) -\lambda }{\sqrt{\left[
\eta _{k}\left( 1+2\Gamma \right) -\lambda \right] ^{2}+\eta _{k}^{2}\Delta
^{2}}}dk\\
\Gamma =& 
-\frac{g}{4\pi}\int_0^\pi\eta _{k}\left[ 1-\frac{\eta _{k}\left( 1+2\Gamma
\right) -\lambda }{\sqrt{\left[ \eta _{k}\left( 1+2\Gamma \right) -\lambda %
\right] ^{2}+\eta _{k}^{2}\Delta ^{2}}}\right] dk
\end{align}
\end{subequations}
with corresponding energy density $e = -\frac{1}{g}\l(\Delta^2 + 4\Gamma(1+\Gamma)\r)$.

For repulsive couplings ($G<0$), there is no mean-field solution ($\Delta =0$%
). The lowest-energy Slater determinant is the ground state of the
noninteracting Fermi gas in one dimension (i.e. our Hamiltonian with $G=0$):
$\ket{\Psi_{\sf{nonint}}}=\prod_{-k_{F}}^{k_{F}}c_{k\uparrow }^{\dagger
}c_{k\downarrow }^{\dagger }\ket0$ for any coupling $G<0$ with energy
\beq
\braket{\Psi_{\sf{nonint}}|H|\Psi_{\sf{nonint}}}
=4\sum_{k=0}^{k_F} \epsilon_k  - G\l(3\sum_{k>0} \epsilon_k^2
+4\l(\sum_{k=0}^{k_F} \epsilon_k\r)^2
\r)
\label{eq:hf_energy}
\eeq
where $\k _{F}$ is the momentum of the highest-energy occupied single-particle state.

\subsection{Ground state energy scaling}
\begin{figure}
    \centering
    \includegraphics{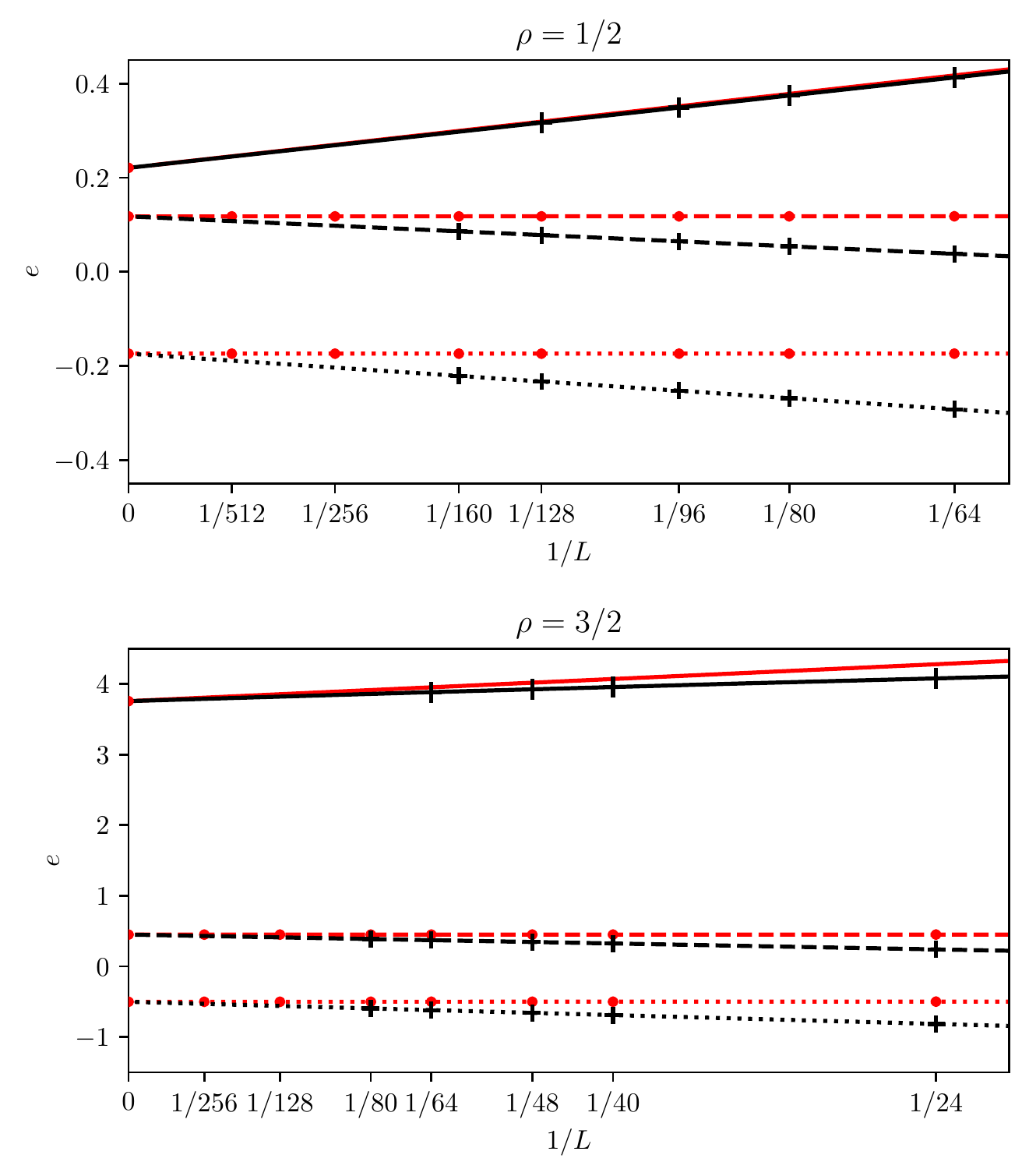}

    \caption{Least-squares fit of energy density $e=\mathcal E/L$ as a function of $1/L$ and couplings $g/g_c = -2$ (solid), $0.8$ (dashes) and $1.6$ (dots) for mean-field (red) and exact (black) results. Crosses and circles mark exact and mean-field solutions, respectively, while for repulsive couplings the energy density was computed continuously from Eq. \eqref{eq:hf_energy}. The circles on the $y$-axis mark the directly-computed thermodynamic values of the mean-field energy densities.}
    \label{fig:scaling}
\end{figure}

We would like to understand which unique properties of these models persist into
the thermodynamic limit (i.e. are present in macroscopic systems). Here, we focus on
the ground state energy density $e_0=\mathcal E_0/L$. In previous work on the SU(2) RG models, it was possible to directly compute the exact energy density in the thermodynamic limit \cite{rombouts10}, but for our purposes, it suffices to plot energy density as a function of $1/L$ and predict the intercept at $1/L=0$. To aid in this prediction, we perform a least-squares fit of this data. These results are plotted in Fig. \ref{fig:scaling} for $\rho=1/2$ in the metallic, trivial superfluid, and nontrivial superfluid phases. Due to the particle-hole symmetry of the model, we expect qualitatively similar results for other densities.

In all cases, the dominant scaling of both the mean-field and exact energy
densities is $1/L$ with corrections of higher order in $1/L$, 
which we fit using the function
\beq
 e(L) = e_\infty + a/L + b/L^2 + c/L^3
\label{eq:rep_fit}
\eeq
where $e_\infty$ is the energy density $e$ as $L\rightarrow\infty$.

For the results in Fig. \ref{fig:scaling}, we find $e_\infty$ consistent with the infinite-size mean-field results up to 
a relative difference of at most $10^{-4}$.
$\e_\infty$ and other fit coefficients for the cases plotted in Fig. \ref{fig:scaling} are presented in Table \ref{tab:fits}.

\begin{table}[h]
\begin{tabular}{l|l|l|l|l|l}
     $g/g_c$ & $\rho$ & $e_\infty$ & $a$ & $b$ & $c$  \\
     \hline
      -2 & $1/2$ & 0.22089 & 12.3 & -0.3 & -6.8 \\
      -2 & $3/2$ & 3.755 & 8.4 & -14.8 & -23.6 \\
      0.8 & $1/2$ & 0.1178 & -5.1 & 1.0 &  3.7\\
      0.8 & $3/2$ & 0.44987 & -5.1 & 0.3 & 4.2\\
      1.2 & $1/2$ & 0.17373 & -7.6 & 0.5 & 5.8\\
      1.2 & $3/2$ & 0.5014 & -7.6 & 1.3 & 7.5
\end{tabular}
\caption{Parameters of the least-squares fit for
exact and mean-field energies plotted in Fig. \ref{fig:scaling}. 
The values for $\e_\infty$ extrapolated from the RG results agree with
the directly-computed mean-field results for all digits included in the
table.}
\label{tab:fits}
\end{table}

\section{Quartets in the SO(5) model}
Quartet states are collective states of four-fermion. 
As mentioned in the main work, at $g=g_c$ and for $N$ divisible by four, 
the Hamiltonian \eqref{eqn:hamiltonian} has a degenerate ground state with eigenstates made of singlet (S=0) quartets. Among them the pure quartet grond state wavefunction
\beq
\ket{\Psi_Q} = \l(\sum_{kk'}\l(T_{1k}\p T_{-1k'}\p + T_{-1k}\p T_{1k'}\p - T_{0k}T_{0k'}\r)\r)^{N/4}\ket 0
\eeq
with total spin $S=0$.
Quartets show up elsewhere in the exact results. For instance, as we can see from \eqref{eqn:NeNw2},
the general, unpolarized 4 fermion ground state also has a quartet form.

What is not as easily determined from exact results is whether or not the ground state for any $N$ away
from $g_c$ supports ground state correlations. Rather than try to measure these correlations directly,
we calculate the quartet gap:
\beq
\Delta_4(N) = \frac{\mathcal E_0(N+2) + \mathcal E_0(N-2) - 2\mathcal E_0(N)}{2}.
\eeq
Analogously to the pair (or charge) gap $\Delta_2(N) = \l(\mathcal E_0(N+1) + \mathcal E_0(N-1) - 2\mathcal E_0(N)\r)/2$, this quantity for $N$ divisible by 4 represents the energy added to the $N$ particle ground state when a quartet is broken into pairs.
The advantage of this compared to other correlation measures is that it can be calculated for large systems using our exact solutions.

It can be easily shown that at $g=0$ (where the ground state energy for $\eta_k=k$ is simply $4\sum_{k=0}^{k_F} k$),
the quartet gap has the value
\beq
\Delta_4^{\sf{nonint}}(N) = \begin{cases}
\frac{\pi}{2L} & \text{$N$ divisible by 4}\\
0 & \text{else}.
\end{cases}
\eeq
This value clearly goes to zero in the limit $L\rightarrow \infty$ and is due to purely kinematic effects. As such, it is important to compare the quartet gaps away from $g=0$ to understand to what extent they are due to legitimate quartet correlations. Additionally,
at $g=g_c$ the quartet gap is zero because the ground state energy for all even values of $N$ is equal. In this case, the zero quartet gap is a consequence of the existence of purely pair-based state that are degenerate with the quartet state.

\begin{figure}
\centering
\includegraphics{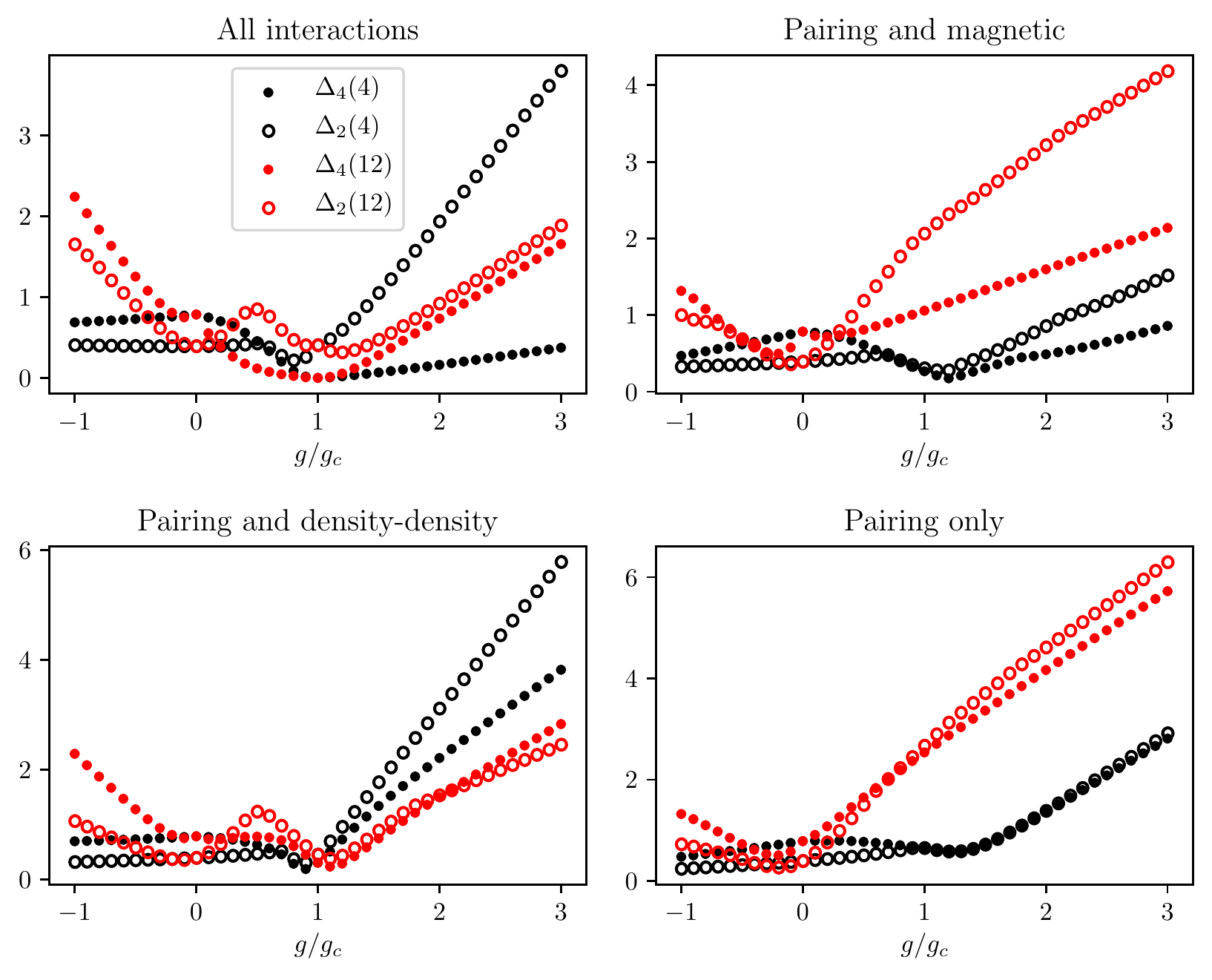}
\caption{Quartet ($\Delta_4(N)$) and pair ($\Delta_2(N)$) gaps for 
$N=4$ and 12, corresponding to $\rho=1/2$ and $3/2$ calculated via exact-diagonalization in a system with $L=8$.}
\label{fig:quartets}
\end{figure}

Quartet gaps, along with pair gaps, are plotted in Fig. \ref{fig:quartets} as a function of $g/g_c$. As can be seen, quartet correlations are most prevalent when the spin-spin interactions are removed from the Hamiltonian, becoming comparable to the pairing correlations for certain densities and couplings. As these modifications break the integrability of the model, these calculations were undertaken using exact-diagonalization in small systems. As such, the thermodynamic behavior of these quantities has yet to be determined.

\section{Highly degenerate flat band limits in pairing Hamiltonians}
\label{app:su2}

The hyperbolic SU(2) Richardson-Gaudin Hamiltonian
\beq
H = \sum_\k \eta_\k^2 c_\k\d c_\k\n - \frac{g}{L} \sum_{\k,\k'}
\eta_\k\n \eta_{\k'}\n
c_{\k}\d c_{-\k}\d c_{-\k'}\n c_{\k'}\n
\label{eqn:su2_hamiltonian}
\eeq
depicts $p$-wave pairing for spinless (or more correctly, 
fully spin-polarized) fermions. This model exhibits topological
superfluidity for $g > 0$ \cite{rombouts10}.
As with the SO(5) RG Hamiltonian, the ground state for $g < 0$
is a metal with atypical excitations. What distinguishes the SU(2)
case is a flat-band limit of the exactly-solvable model,
initially introduced to study fractional Quantum
Hall physics \cite{ortiz13}:
\beq
H_\infty = \lim_{g\rightarrow-\infty}\frac{1}{g}H
= \sum_{\k,\k'}\eta_\k^{\;}  \eta_{\k'}^{\;} 
c_{\k}\d c_{-\k}\d c_{-\k'}^{\;}  c_{\k'}^{\;} .
\label{eqn:su2_flatband}
\eeq
Here, solutions to the SU(2) RG equations
may be classified by the number of singular pairons. All solutions with
all pairons finite give zero energy in the limit due to the factor of 
$1/G$. The only other possible solutions have
finite energy with precisely one pairon diverging.
By counting the ways of arranging the divergent pairons
and subtracting this number from the total number of available
states, the degeneracy of the ground state is found to be
\beq
d^{\nu=0}_0=\max\l(1, {L/2 \choose N/2} - {L/2 \choose N/2-1}\r)
\eeq
for a system with $N/2$ fermion pairs on $L$ sites \cite{ortiz13}.
Extrapolating to all seniorities gives total ground state degeneracy
\beq
d_0^{\sf{total}}=\max\l(1, {L \choose N} - {L \choose N-2}\r).
\eeq
As a fraction of the total Hilbert space dimension
$|\mathcal H| = {L\choose N}$, this is
\beq
\frac{d^{\sf{total}}_0}{|\mathcal H|}
=\max\l(0, \frac{1-2\rho}{(1-\rho)^2}\r)
\eeq
in the thermodynamic limit.

\begin{figure}
\centering
\includegraphics{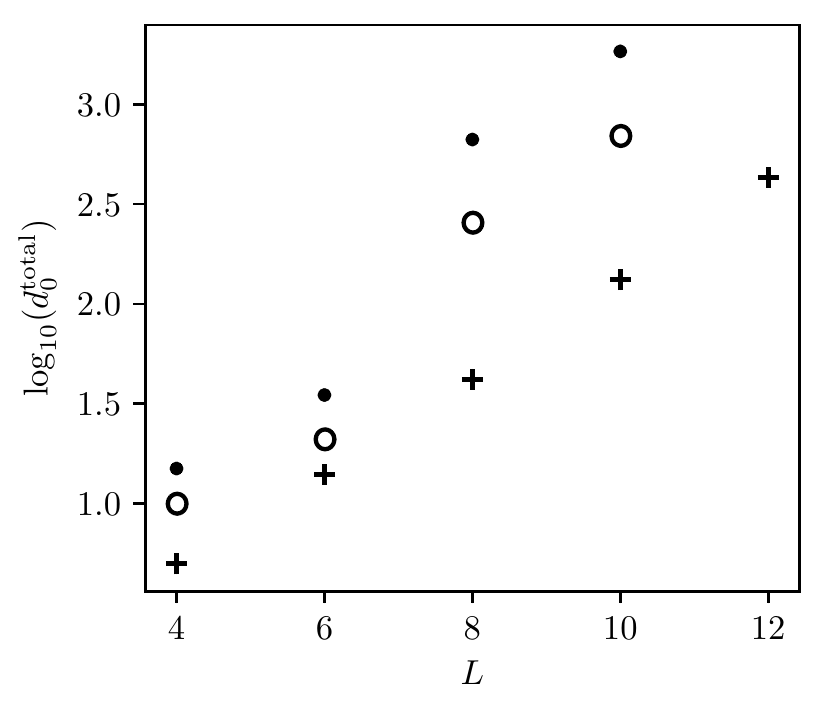}
\caption{Logarithmic plot showing ground state degeneracy
for the Hamiltonian \eqref{eqn:hamiltonian2}
in the strong coupling limit
with pairing and spin interactions (open circles)
and
only pairing interactions (dots). As a comparison,
the degeneracy is also plotted for the SU(2)
Hamiltonian \eqref{eqn:su2_hamiltonian} in the same
limit (crosses). All degeneracies are calcualated at $\rho = 1/2$.}
\label{fig:degens}
\end{figure}

A relative of our exactly-solvable Hamiltonian \eqref{eqn:hamiltonian},
\begin{align}
\begin{split}
H' =& \sum_{\k} \e_\k\n N_\k \n
-G_T \sum_{\k,\k'}\eta_\k\n \eta_{\k'}\n
\vec T_\k\p \cdot \vec T_{\k'}\m
-G_S \sum_{\k,\k'}\eta_\k\n \eta_{\k'}\n S_\k\p S_{\k'}\m,
\end{split}\label{eqn:hamiltonian2}
\end{align}
also exhibits an exponentially degenerate, zero energy
ground state in the limit $\epsilon_\k \rightarrow 0$ with 
$\eta_\k$ finite, although the precise formula for this degeneracy cannot be derived from the RG equations as was done for the SU(2) model \cite{ortiz13}. The number of ground states is plotted as a function of system size in Fig. \ref{fig:degens}.
In this zero kinetic-energy limit with $G_S=0$,
this is the Haldane-Rezayi Hamiltonian \cite{haldane88}.
Near this limit, this Hamiltonian has an
extremely high density of states close to the ground state,
creating a nontrivial excitation spectrum
and non-Fermi liquid behavior. For lower couplings, 
both the exactly-solvable SO(5) Hamiltonian 
\eqref{eqn:hamiltonian} and Hamiltonian
\eqref{eqn:hamiltonian2} have Fermi-liquid-like behavior.

\begin{figure}
\centering
\includegraphics{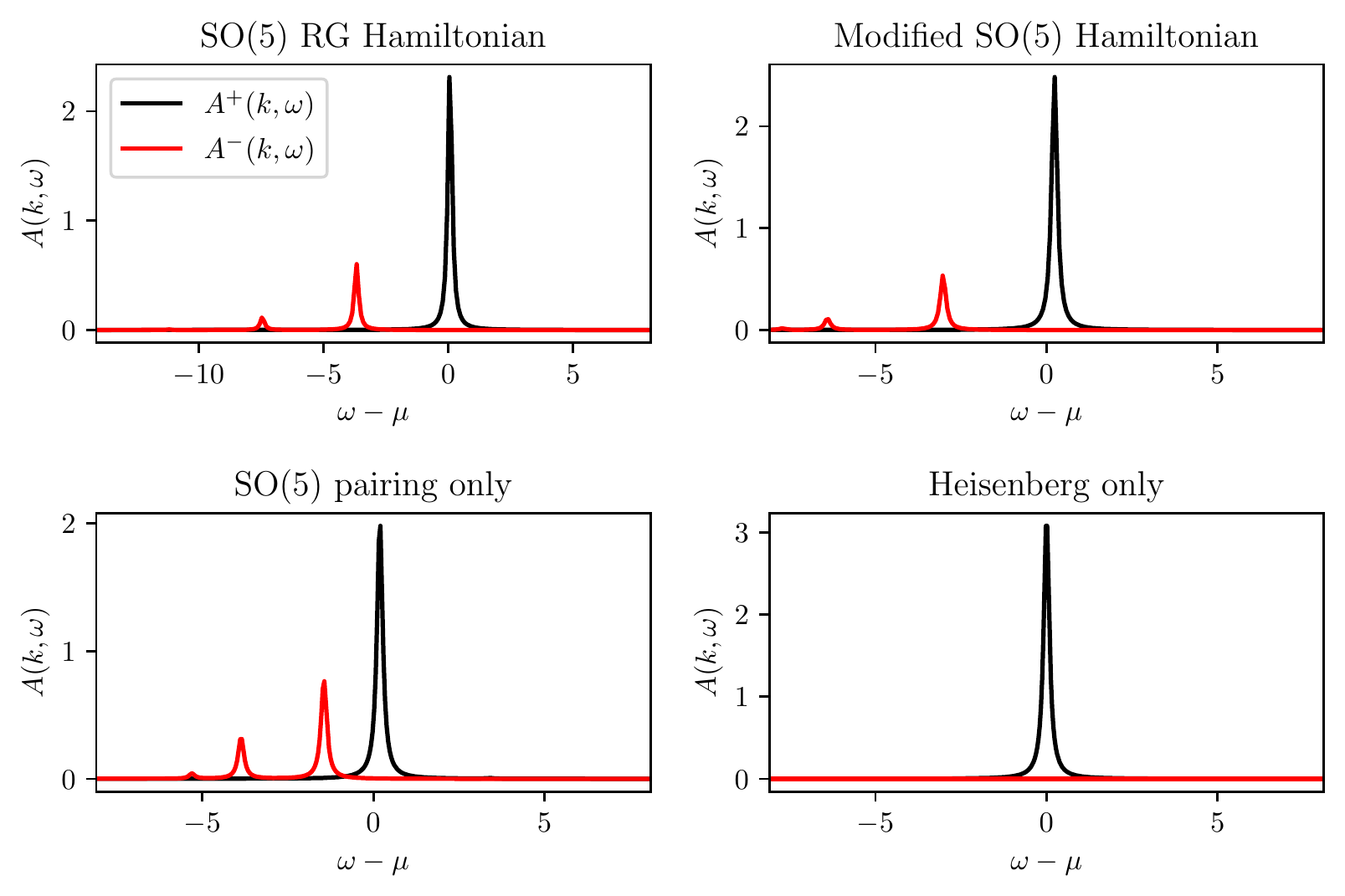}
\includegraphics{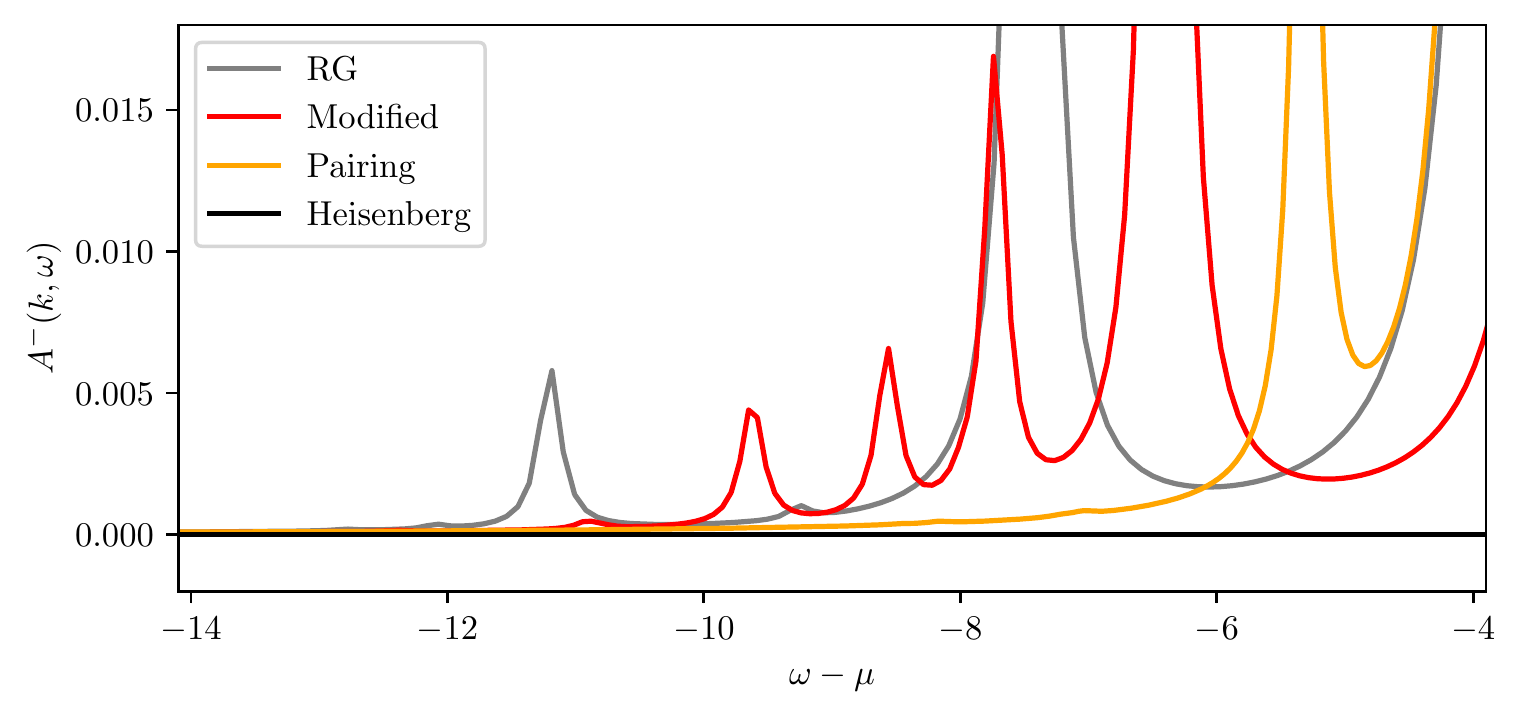}

\caption{Spectral function for various SO(5) 
Hamiltonians in a system with $L=8$ at three-quarter filling ($\rho = 3/2$) and 
$g/g_c = -1$.
The spectral function \eqref{eqn:spectral} is calculated for the lowest-energy unoccupied
mode, and with $\epsilon=0.2$ to smooth out
the delta peaks.
The modified Hamiltonian is Eq. \eqref{eqn:hamiltonian2}.
In all cases, $A^+(k,\omega)$ has a large
peak at or near the Fermi energy $\mu=\frac{1}{2}\l(\mathcal E_0(N+1) - \mathcal E_0(N-1)\r)$.
In the cases of kinetic energy plus Heisenberg interactions and (not pictured) purely kinetic or kinetic plus density-density, this is the only peak in the spectral function, indicating single-particle excitations.
In the other cases, a number of peaks appear
in $A^-(k,\omega)$, as discussed in the text.
The lower plot shows just $A\m(k,\omega)$ for these
same systems over a smaller range in $\omega-\mu$ to emphasize features not visible 
in the upper plots.
}
\label{fig:spectral}
\end{figure}

We can contrast this behavior with two limits of the exactly-solvable
SO(5) Hamiltonian. First, we copy the SU(2) case and take the
coupling $G \rightarrow -\infty$, achievable by setting
$q=-1/\sum_\k \eta_\k$.
In this limit, there is the same finite degeneracy as seen at lower
repulsive couplings.
The absence of exponential degeneracy is because the SO(5) interactions
contain single-particle terms
that produce a nonzero effective kinetic energy. Pulling some
of these terms out of the interactions, the Hamiltonian is
\begin{eqnarray}
H &=& \sum_\k \l(\eta_\k-\frac{3G}{2}\eta_\k^2\r)N_\k\n
-G \sum_{\k,\k'}\eta_\k\n \eta_{\k'}\n \l(
2\vec T_{\k}\p \cdot \vec T_{\k'}\m 
+ \vec S_\k \cdot \vec S_{\k'} + \frac{1}{4}N_\k\n N_{\k'}\n
\r),
\end{eqnarray}
with additional single-particle terms from the commutators of
$S_\k\p S_{\k'}\m$ and diagonal entries of the $S_\k\z S_{\k'}\z$ and
$N_\k\n N_{\k'}\n$ terms. As these scale with $G$, it is impossible to have the many-body interactions have more than a 
$\mathcal O(1)$ strength relative to the single-particle terms.

Unlike the SU(2) case \cite{rombouts10, ortiz13},
these single-particle terms cannot be absorbed into the kinetic energy by rescaling and defining
a new coupling, since they have coefficients $\eta_\k^2$ rather than
$\eta_\k$. The rational SU(2) Hamiltonian with density-density
interactions studied in Ref. \cite{stouten19} behaves similarly
to the SO(5) case in the infinite-coupling limit, for the same
reasons.
The Hamiltonian
\eqref{eqn:hamiltonian2} is arrived at by simply
removing all these terms.

At the critical point $G_c$, the integrable SO(5) Hamiltonian is proportional to $\sum_{kk'}\eta_k \eta_{k'}\vec{\mathcal T}_k \cdot \vec{\mathcal T}_{k'}$, again superficially similar to 
Eq. \eqref{eqn:su2_flatband}. As with the $G \rightarrow - \infty$
limit just examined, single-particle terms appearing
in the interaction spoil the exponential degeneracy by creating
an effective kinetic energy. Although there is a ground
state degeneracy at this point,
it scales only linearly with the system size, and thus represents
a vanishing fraction of the full Hilbert space in the thermodynamic
limit.

To understand the nature of excitations in these models,
we make use of the spectral function $A(k,\sigma,\omega) = A^+(k, \sigma,\omega) + A^-(k,\sigma,\omega)$,
here defined as
\beq
A^\pm (k,\sigma,\omega) = \frac{1}{\pi}\text{Im} \sum_\alpha
\frac{\bra{\Psi_\alpha}c_{k\sigma}^\pm \ket{\Psi_0}}
{\mathcal E_0 - \mathcal E_\alpha \pm \omega - i\epsilon},
\quad
c^+_{k\sigma}=c\d_{k\sigma} = \l(c\m_{k\sigma}\r)\d,
\label{eqn:spectral}
\eeq
where $\alpha$ indexes all eigenstates $\ket{\Psi_\alpha(N\pm 1)}$ of the Hamiltonian with energies $\mathcal E_\alpha(N\pm 1)$ and $\epsilon$
is a positive, infinitesimal number. Due to symmetries of the systems we are concerned with, we will take $\sigma=\uparrow$ for all of the following ($A(k,\omega) = A(k, \uparrow, \omega)$. We plot these functions for various modifications of the SO(5) Hamiltonian in Fig. \ref{fig:spectral}.

A single peak in $A(k,\omega)$ for $k$ fixed indicates single-particle excitations of the $N$-particle ground state are close to eigenstates of the Hamiltonian, allowing for accurate quasiparticle descriptions of these eigenstates. This peak is located at a value of $\omega$ equal to the energy difference between the $N$ and $N\pm 1$-particle states, which when $k=k_F$ (the Fermi momentum), corresponds to the Fermi energy $\mu$. This is emblematic of Fermi liquid states, where the infinitesimally thin peak of a noninteracting Fermi gas acquires a finite width and therefore gives the quasiparticles a finite lifetime. Such behavior is seen in SO(5) Hamiltonians with kinetic energy and only Heisenberg or (not pictured) density-density interactions.
As (repulsive) pairing interactions are added, a number of additional peaks are formed away from $\omega=\mu$. As plotted, we choose $k$ slightly above the Fermi momentum so the main peak is in $A^+(k,\omega)$ (as the additional fermion is created/annihilated above the Fermi level), while the additional behavior is seen in $A^-(k,\omega)$.